\documentclass[%
reprint,
superscriptaddress,
showkeys,
amsmath,
amssymb,
aps,
rmp,
floatfix,
]{revtex4-2}
\usepackage{xcolor}
\usepackage{graphicx}
\usepackage{dcolumn}
\usepackage{bm}
\usepackage{hyperref}
\usepackage[mathlines]{lineno}

\usepackage{stmaryrd} 
\usepackage{soul} 
\usepackage{placeins} 
\usepackage{subcaption} 
\usepackage{enumerate} 
\usepackage{yhmath} 

\newcommand{\lb}{\llbracket}
\newcommand{\rb}{\rrbracket}
\DeclareMathOperator{\diag}{diag}
\DeclareMathOperator{\Cov}{cov}

\begin{document}

\title{When is cross impact relevant?}

\author{Victor Le Coz}
\email{victor.lecoz@quant.global}
\affiliation{Chair of Econophysics and Complex Systems, \'Ecole polytechnique, 91128 Palaiseau Cedex, France}
\affiliation{LadHyX UMR CNRS 7646, \'Ecole polytechnique, 91128 Palaiseau Cedex, France}
\affiliation{Quant AI lab, 29 Rue de Choiseul, 75002 Paris, France}

\author{Iacopo Mastromatteo}
\email{iacopo.mastromatteo@cfm.com}
\affiliation{Capital Fund Management, 23 Rue de l’Universit\'e, 75007 Paris, France}

\author{Damien Challet}
\email{damien.challet@centralesupelec.fr}
\affiliation{Laboratoire de Math\'ematiques et Informatique pour la Complexit\'e et les Syst\`emes, CentraleSupélec, Universit\'e Paris-Saclay, 91192 Gif-sur-Yvette Cedex, France}

\author{Michael Benzaquen}
\email{michael.benzaquen@polytechnique.edu}
\affiliation{Chair of Econophysics and Complex Systems, \'Ecole polytechnique, 91128 Palaiseau Cedex, France}
\affiliation{LadHyX UMR CNRS 7646, \'Ecole polytechnique, 91128 Palaiseau Cedex, France}
\affiliation{Capital Fund Management, 23 Rue de l’Universit\'e, 75007 Paris, France}

\date{\today}

\begin{abstract}
Trading pressure from one asset can move the price of another, a phenomenon referred to as \textit{cross impact}. Using tick-by-tick data spanning $5$ years for $500$ assets listed in the United States, we identify the features that make cross-impact relevant to explain the variance of price returns. We show that price formation occurs endogenously within highly liquid assets. Then, trades in these assets influence the prices of less liquid correlated products, with an impact velocity constrained by their minimum trading frequency. We investigate the implications of such multidimensional price formation mechanism on interest rate markets. We find that the 10-year bond future serves as the primary liquidity reservoir, influencing the prices of cash bonds and futures contracts within the interest rate curve. Such behaviour challenges the validity of the theory in Financial Economics that regards long-term rates as agents anticipations of future short term rates.
\end{abstract}


\maketitle


\section{Introduction} \label{Introduction}

According to standard economic theory, the price of an asset should integrate all publicly available information regarding its fundamental value. In practice, price formation occurs through a trading system that mechanically forces information flow into prices via the order flow of market participants. This well-established phenomenon is known as \textit{price impact}.

An early model of price impact was proposed by \citet{Kyle-1985}, who assumed a linear dependence between absolute price differences and signed traded volumes. Further work established that the price impact of large (split) trades universally follows a square-root law of the traded volume \citep{Loeb-1983,PlerouEtAl-2004,AlmgrenEtAl-2005a,KissellMalamut-2005,MoroEtAl-2009,TothEtAl-2011,MastromatteoEtAl-2014,DonierBonart-2015,ZarinelliEtAl-2015,KyleObizhaeva-2023,BacryEtAl-2015,TothEtAl-2017,BouchaudEtAl-2018}. Yet, the price impact of a single anonymous market order is a much weaker concave (almost constant) function of its volume when the latter is adequately normalized by the available liquidity in the order book \citep{Hasbrouck-1991,ChenEtAl-2002,LilloEtAl-2003,PottersBouchaud-2003,Zhou-2012,GomberEtAl-2015,BouchaudEtAl-2018}. This behavior is due to the \textit{selective liquidity taking} effect \citep{TarantoEtAl-2014,BouchaudEtAl-2018}: most of the large market order arrivals happen when there is a large volume available at the opposite-side best quote, specifically trying to avoid moving the mid-price. To overcome this effect, impact is often measured over a coarse-grained time scale~$\tau$, by aggregating trades into a \textit{signed order flow imbalance}. This method involves calculating the signed sum of the volumes of all trades within a time window of length~$\tau$, while observing price changes during the same interval. Within this framework, the magnitude of price impact crosses over from a linear to a concave behavior, as the signed order flow increases \citep{KempfKorn-1999,PlerouEtAl-2002,EvansLyons-2002,ChordiaSubrahmanyam-2004,GabaixEtAl-2006,Hopman-2007,PatzeltBouchaud-2017,PatzeltBouchaud-2018,BouchaudEtAl-2018}. In addition, the aggregated price impact is a concave function of the time scale~$\tau$ chosen for the aggregation \citep{BouchaudEtAl-2018}. Finally, in order to conciliate the long term positive auto-correlation of trades with the independence of price increments, \citet{BouchaudEtAl-2004} established that price impact must be \textit{transient}. This assumption means that the magnitude of the price-impact of a trade decreases across time. This hypothesis was corroborated in the following years \citep{BouchaudEtAl-2006,Hopman-2007,BouchaudEtAl-2009a,Gatheral-2010,GatheralSchied-2013,AlfonsiEtAl-2016,GarleanuPedersen-2016,TothEtAl-2017,TarantoEtAl-2018, EkrenMuhle-Karbe-2019}.

A more subtle effect is that trading pressure from one asset can move the price of another. This effect, which is referred to as \textit{cross-impact}, was studied initially by \citet{HasbrouckSeppi-2001} and later in \citet{ChordiaEtAl-2001, EvansLyons-2001, HarfordKaul-2005, PasquarielloVega-2007, AndradeEtAl-2008, Tookes-2008, PasquarielloVega-2015, WangGuhr-2017,BenzaquenEtAl-2017, SchneiderLillo-2019, TomasEtAl-2021, TomasEtAl-2022, BrigoEtAl-2022}.

The simplest cross-impact models posit a linear relationship between signed trading volumes and prices variations in time windows of length~$\tau$ (the binning frequency) \citep{HasbrouckSeppi-2001,HarfordKaul-2005,PasquarielloVega-2007,PasquarielloVega-2015,TomasEtAl-2021,TomasEtAl-2022}. While the time decay of the transient impact model was studied for bonds \citep{SchneiderLillo-2019,Schneider-2019} and stocks \citep{Wang-2017}, the time scale maximizing the accuracy of linear cross-impact models has not yet been documented. Moreover, this optimal time scale is an indicator of the speed of information transmission among assets, which has not been studied extensively, although \citet{ZumbachLynch-2001,LynchZumbach-2003,CordiEtAl-2021} inferred typical time scales of market reactions of the volatility process. In addition, \citet{TomasEtAl-2022,RosenbaumTomas-2022,CordoniEtAl-2022} link the magnitude of cross-impact to asset liquidity and to the correlation among assets.

Here, we quantitatively characterize the circumstances under which a model with cross-impact over-performs one that does not include impact across assets. Additionally, we identify the time scales that maximize the accuracy of linear cross-impact models. Our study includes an introduction to the linear cross-impact modeling framework and a methodology to evaluate the factors influencing cross-impact's relevance in explaining price return variance. The results are organized according to the studied features: the bin size, the trading frequency, the correlation among assets, and the liquidity. In the final section, we provide applications of these findings to the interest rate curve.

\section{Notations}
The set of real-valued square matrices of dimension~$n$ is denoted by $\mathcal{M}_n(\mathbb{R})$, the set of orthogonal matrices by $\mathcal{O}_n$, the set of real symmetric positive semi-definite matrices by $\mathcal{S}^{+}_n(\mathbb{R})$, and the set of real symmetric positive definite matrices by $\mathcal{S}^{++}_n(\mathbb{R})$. Given a matrix~$A$ in~$\mathcal{M}_n(\mathbb{R})$, $A^\top$ denotes its transpose. Given~$A$ in~$\mathcal{S}^{+}_n(\mathbb{R})$, we write $A^{1/2}$ for a matrix such that $A^{1/2}(A^{1/2})^\top = A$, and $\sqrt{A}$ for the matrix square root: the unique positive semi-definite symmetric matrix such that $(\sqrt{A})^2 = A$. We also write $\diag(A)$ for the vector in~$\mathbb{R}^n$ formed by the diagonal items of $A$. Finally, given a vector~$v$ in~$\mathbb{R}^n$, we denote the components of $v$ by $(v_1, \cdots, v_n)$, and the diagonal matrix whose components are the components of $v$ by $\diag(v)$. Table~\ref{tab:notations} in appendix~\ref{Notations} provides the complete list of the notations used in this study.

\section{Modeling framework}
To relate trades to prices, we observe the mid-prices and market orders of $n$ different assets, both binned at a regular time interval of length~$\tau$ seconds. We denote by $p_{t,i}$ the opening price of asset~$i$ in the time window~$[t, t + \tau ]$ and by $p_t = (p_{t,1}, \cdots, p_{t,n} )$ the vector of asset prices at opening. We define $q_{t,i}$ as the net market order flow traded during the time window~$[t, t + \tau ]$. This is calculated by taking the sum of the volumes of all trades during that time period, with buy trades counted as positive and sell trades counted as negative. Hence, $q_t = (q_{t,1}, \cdots, q_{t,n} )$ is the vector of the net traded order flows.

Following the approach proposed by \citet{TomasEtAl-2022}, we study the relationship between the time series of net order flows $\{q_0, q_{\tau}, \cdots \}$ and the time series of prices $\{p_0, p_{\tau}, \cdots \}$, under the two following assumptions:
\begin{itemize}
\item prices variations~$\Delta p_t := p_{t+\tau} - p_t$ and order flow imbalances~$q_t$ are linearly related, i.e.,
\begin{equation}
    \Delta p_t = \Lambda_t q_t + \eta_t, 
\end{equation}
where the $n \times n$~matrix~$\Lambda_t$ is the cross-impact matrix and $\eta_t = (\eta_{t,1}, \cdots, \eta_{t,n} )$ is a vector of zero-mean random variables representing exogenous noise;

\item the cross-impact matrix~$\Lambda_t$ is a function of the form:
\begin{equation}
\label{eq:prices_dynamics}
\Lambda_t = \Lambda_t(\Sigma_t, \Omega_t, R_t),
\end{equation}
where $\Lambda_t: \mathcal{S}^{+}_n(\mathbb{R}) \times \mathcal{S}^{+}_n(\mathbb{R}) \times \mathcal{M}_n(\mathbb{R}) \to \mathcal{M}_n(\mathbb{R})$ is called a cross-impact model, $\Sigma_t := \Cov(\Delta p_t)$ is the price variations covariance matrix, $\Omega_t := \Cov(q_t)$ is the order flows covariance matrix, and $R_t = \mathbb{E}\left[ (\Delta p_t - \mathbb{E}\left[ \Delta p_t \right])(q_t - \mathbb{E}\left[ q_t \right])^\top\right]$ is the response matrix.
\end{itemize}

We also define the price variations volatility by
\begin{equation}
    \sigma_t := (\sqrt{\Sigma_{t,11}}, \cdots, \sqrt{\Sigma_{t,nn}}),
\end{equation}

and the signed order flows volatility by
\begin{equation}
    \omega_t := (\sqrt{\Omega_{t,11}}, \cdots, \sqrt{\Omega_{t,nn}}).
\end{equation}

Finally, for a given asset~$i$, we define the the average across time of its price variation volatility by 
\begin{equation}
    \bar{\sigma}_i := \langle \sigma_{t,i} \rangle, 
\end{equation}
and the average across time of its signed order flow volatility by
\begin{equation}
    \bar{\omega}_i := \langle \omega_{t,i} \rangle.
\end{equation}

\subsection{Definition of the models}
Let $Y$ denote a scalar called the \textit{Y-ratio}. We study the following three cross-impact models:
\begin{itemize}

\item the diagonal model, defined by
\begin{equation}
    \Lambda_{\text{diag}}(\Sigma, \Omega, R) := Y \diag(R)\diag(\Omega^{-1}),
\end{equation}
which is the limit case where the cross-sectional impact is set to zero;

\item the Maximum Likelihood model (\textit{ML model} in the following sections), defined by
\begin{equation}
    \Lambda_{\text{ML}}(\Sigma, \Omega, R) := Y R \Omega^{-1};
\end{equation}

\item and the Kyle model, defined by
\begin{equation}
    \Lambda_{\text{Kyle}}(\Sigma, \Omega, R) := Y (\Omega^{-1/2})^\top \sqrt{(\Omega^{1/2})^\top \Sigma\Omega^{1/2}}\ \Omega^{-1/2}.
\end{equation}

\end{itemize}
The Y-ratio is a re-scaling adjustment parameter, estimated by minimizing, across all assets, the squared errors between the price variations predicted by the model and the realized prices. These models were investigated in several of the previously mentioned  publications. Noteworthy, the diagonal model was studied in \citep{KempfKorn-1999,PlerouEtAl-2002,EvansLyons-2002,ChordiaSubrahmanyam-2004,GabaixEtAl-2006,Hopman-2007,PatzeltBouchaud-2017,PatzeltBouchaud-2018,BouchaudEtAl-2018}. The Maximum Likelihood model was investigated in \citet{HasbrouckSeppi-2001,HarfordKaul-2005,PasquarielloVega-2007,PasquarielloVega-2015,TomasEtAl-2021,TomasEtAl-2022}. The multidimensional Kyle model was introduced by \citet{delMolinoEtAl-2018} and further Examined by \citet{TomasEtAl-2022}.

It is important to note that the diagonal model can be defined by 
\begin{equation}
    \Lambda_{\text{diag}} := \diag(\lambda_{\text{diag}}),
\end{equation}
where the vector $\lambda_{\text{diag}} = (\lambda_1, \cdots, \lambda_n)$ is defined by a set of linear equations:
\begin{equation}
\forall i \in \lb 1,n \rb, \; \Delta p_{t,i} = \lambda_i q_{t,i} + \eta_{t,i}.
\end{equation}
 This means that the diagonal model assumes that each asset~$i$ has its own unique relationship between price increments and order flows, as captured by the coefficient~$\lambda_i$.

The comparison between the last two models and the first one will allow us to distinguish among the portion of cross-impact that is explained by order flow commonality (which the diagonal model can capture) and the contributions that cannot be explained by this effect, thus requiring models such as ML and Kyle.

\subsection{Properties of the models} \label{Models' properties}

As demonstrated by \citet{TomasEtAl-2022}, the previously defined models satisfy a list of properties that characterize their behavior. These properties are recalled below.
\begin{enumerate}
\item {\em Symmetry} properties aim at ensuring that the cross-impact model behaves in a controlled manner under financially-grounded transformations of its variables~$\Sigma_t, \; \Omega_t, \; R_t$. The Kyle and ML models both adapt to (i) a re-ordering of the considered assets (\textit{permutation invariance}), (ii) a change of currency (\textit{cash invariance}) or (iii) volume units (\textit{split invariance}) and, (iv) a change of basis in the asset space (\textit{rotational invariance}). In contrast, the diagonal model crucially misses property (iv), as it regards the physical space of assets as a privileged basis for the description.
\item {\em Non-arbitrage} properties aim at ensuring the absence of arbitrage in the sense of \cite{Gatheral-2010}, i.e., round-trip trading strategies with positive average profit. Both the diagonal and Kyle model prevent (i) static arbitrage over a single-period (thanks to their \textit{positive semi-definiteness}) and (ii) dynamic arbitrage over multi-period. Yet, the ML model does not satisfy any of these non-arbitrage properties.
\item {\em Fragmentation} properties aim at ensuring the equality of the price impacts generated from traded volumes of the same assets on fragmented markets (e.g. US stocks are traded on several venues). This property is satisfied by both the Kyle and ML models but is trivially violated by the diagonal model.
\item {\em Stability} properties aim at ensuring the impossibility to manipulate the price of liquid products using illiquid instruments. This property is satisfied by the Kyle model and the diagonal model, but not by the ML model.
\end{enumerate}

\section{Methodology}

\subsection{Estimation method} \label{Estimation method}

We use tick-by-tick trades and quotes for $500$ assets quoted in limit order books in the United States. Our sample includes stocks, bonds, futures on bonds and futures on stock indexes. Unless otherwise specified, our data set covers the $2017-2022$ period. For a given year, we consider the data from the preceding year as in-sample data, while the data from the current year is designated as out-of-sample data. We then aggregate in and out-of-sample results over all years.

A significant portion of our analysis involves the selection of pairs of assets from the pool of the $500$ assets in our sample. For this purpose, we select $20,000$ pairs of assets per year. These pairs are chosen from the pool of $\approx$120,000 possible combinations among the 500 assets, aiming for a uniform coverage of existing correlations. Specifically, we categorize all potential pairs based on their correlation levels into 50,000 equally-sized correlation buckets. For instance, the first bucket encompasses pairs with correlations between $0\%$ and $0.002\%$. Subsequently, we opt for the first pair within each of these buckets, resulting in the selection of 20,000 pairs. Our analysis is then aggregated over a five-year period, yielding a total of 100,000 year-pair combinations.

To overcome the conditional heteroskedasticity of price variations and signed order flows, we use a daily estimator of their volatility. Let $\{t_1, \cdots, t_K\} \in \mathbb{R}^K$ denote the $K$~business days of a year. For each day~$t_k$, the estimators of the price increments volatility and of the signed order flows volatility are defined by
\begin{equation}
\begin{aligned}
\widehat{\sigma}_{t_k} := \left(\sqrt{\langle {\Delta p_{t,1}}^2 \rangle_{t_k}}, \cdots, \sqrt{\langle{\Delta p_{t,n}}^2\rangle_{t_k}}\right), \\
\widehat{\omega}_{t_k} := \left(\sqrt{\langle {q_{t,1}}^2 \rangle_{t_k}}, \cdots, \sqrt{\langle{q_{t,n}}^2\rangle_{t_k}}\right), \\
\end{aligned}
\end{equation}
respectively, where the average $\langle.\rangle_{t_k}$ is computed using data on the day~$t_k$. We assume the correlation matrices $\rho_{\Delta p}=\diag({\sigma}_t)^{-1} \Sigma_t \diag({\sigma}_{t})^{-1}$ and $\rho_q=\diag({\omega}_t)^{-1} \Omega_t \diag({\omega}_{t})^{-1}$, as well as the normalized response matrix $\rho_{\Delta p,q} =\diag({\sigma}_t)^{-1} R_t \diag({\omega}_{t})^{-1}$, are stationary. Let $\widehat{\rho}_{\Delta p}$, $\widehat{\rho}_{q}$ and $\widehat{\rho}_{\Delta p,q}$ denote their respective long period estimators, i.e.
\begin{equation}
\begin{aligned}
   \widehat{\rho}_{\Delta p} = \langle \rho_{\Delta p} \rangle, \\
    \widehat{\rho}_{q} = \langle \rho_q \rangle, \\
    \widehat{\rho}_{\Delta p,q} = \langle \rho_{\Delta p,q} \rangle, \\
\end{aligned}
\end{equation}
respectively, where the average $\langle.\rangle$ is computed using one year of data. The covariance of any variables $x$ and $y$ can be expressed as the product of their correlation and each of their respective volatilities. Thus, on the day $t_k$, the estimated covariance and response matrices $\widehat{\Sigma}_{t_k}$, $\widehat{\Omega}_{t_k}$ and $\widehat{R}_{t_k}$ are obtained by
\begin{equation}
\begin{aligned}
    \widehat{\Sigma}_{t_k} = \diag(\widehat{\sigma}_{t_k}) \widehat{\rho}_{\Delta p} \diag(\widehat{\sigma}_{t_k}), \\
    \widehat{\Omega}_{t_k} = \diag(\widehat{\omega}_{t_k}) \widehat{\rho}_{q} \diag(\widehat{\omega}_{t_k}), \\
    \widehat{R}_{t_k} = \diag(\widehat{\sigma}_{t_k}) \widehat{\rho}_{\Delta p,q} \diag(\widehat{\omega}_{t_k}), \\
\end{aligned}
\end{equation}
respectively.

\subsection{Metrics definition}

\subsubsection{Goodness-of-fit}

For a given cross-impact model~$\Lambda_t$, the predicted price change for the time window~$\left[t, t + \tau \right]$ due to the order flow imbalance~$q_t$ is defined as
\begin{equation}
    \widehat{\Delta p}_t := \Lambda_t(\widehat{\Sigma}_t, \widehat{\Omega}_t, \widehat{R}_t) q_t.
\end{equation}

To evaluate the goodness-of-fit of the cross-impact model~$\Lambda$, we compare the predicted price changes~$\widehat{\Delta p}_t$ to the realized price changes~$\Delta p_t$. For this purpose, we use a performance indicator parameterized by a symmetric, positive definite matrix~$M \in \mathcal{S}^{+}_n(\mathbb{R}), \; M \neq 0$. Let $\{t_1, \cdots, t_N\} \in \mathbb{R}^N$ denote $N$ sample times, $\{\Delta p_{t_1}, \cdots, \Delta p_{t_N}\}$ be a realization of the price process, and $\{\widehat{\Delta p}_{t_1}, \cdots, \widehat{\Delta p}_{t_N}\}$ denote the corresponding series of predictions from the model. The $M$-weighted generalized $\mathcal{R}^2(M)$ is defined as
\begin{equation}
    \mathcal{R}^2(M) := 1 - \frac{\Sigma_{i=1}^N\left(\Delta p_{t_i} - \widehat{\Delta p}_{t_i}\right)^\top M \left(\Delta p_{t_i} - \widehat{\Delta p}_{t_i}\right)}{\Sigma_{i=1}^N\Delta p_{t_i}^\top M \Delta p_{t_i} }.
\end{equation}

The closer this score is to one, the better is the fit to the actual prices. To highlight different sources of error, different choices of $M$ can be considered:
\begin{itemize}
\item $M = I_\sigma := {\diag(\langle\sigma_t^2\rangle)}^{-1}$, to account for errors relative to the typical deviation of the assets considered. This type of error is relevant for strategies predicting idiosyncratic moves of the constituents of the basket, rather than strategies betting on correlated market moves.
\item $M = I_{\sigma_i} := {\diag(\langle\sigma_{t,i}^2)\rangle}^{-1}$, to account for errors of a single asset~$i$.
\item $M = \langle\Sigma_t\rangle^{-1}$, to consider how well the model predicts the individual modes of the return covariance matrix. This would be the relevant error measure for strategies that place a constant amount of risk on the modes of the correlation matrix, leveraging up combinations of products with low volatility and scaling down market direction that exhibit large fluctuations.
\end{itemize}

Within the following sections, we mainly study the cross-impact goodness-of-fit for pairs~$(i,j)$ of assets. In these cases, unless stated otherwise, we calculate $\mathcal{R}^2(I_{\sigma_i})$ to measure solely the errors on the first asset~$i$, the predicted asset, as a function of the characteristics of the second asset~$j$, the explanatory asset.

Additionally, we define a second indicator to determine the extent to which the goodness-of-fit results from cross-sectional information. We define $\Delta \mathcal{R}^2(M)$, the accuracy increase from the cross sectional model as
\begin{equation}
    \Delta \mathcal{R}^2(M) := \mathcal{R}^2(M) - \mathcal{R}_{\text{diag}}^2(M),
\end{equation}
where $\mathcal{R}_{\text{diag}}^2(M)$ is the $M$-weighted generalized $\mathcal{R}^2(M)$ obtained from the degenerated model without cross-sectional impact $\Lambda_{\text{diag}}$.

Tests that confirm the statistical significance of the $\mathcal{R}^2$ in the case of a single asset are reported in the appendix~\ref{Statistical significance of the Goodness-of-fit}. Nevertheless, appendix~\ref{Auto-correlation structure and comparison with the propagator model} demonstrates that the auto-correlation of signed order flows invalidates the linear framework used in this study. Using of a more accurate \textit{propagator model} \citep{BouchaudEtAl-2006,Bouchaud-2009, AlfonsiEtAl-2016, BenzaquenEtAl-2017, BouchaudEtAl-2018, SchneiderLillo-2019} would yield only marginal improvements in the goodness-of-fit \citep{TomasEtAl-2022} but it would impede conducting this study at the same scale across time and assets.

\subsubsection{Definition of the assets characteristics}

Several assets characteristics are investigated in this study:
\begin{itemize}
    \item the \textit{bin size}~$\tau$;
    \item the \textit{trading frequency}~$f$, defined by the number of trades per second;
    \item the \textit{price increments correlation}~$\rho_{ij}$ between the assets~$i$ and~$j$;
    \item the \textit{liquidity}, defined by the risk of profit or loss in monetary units $\bar{\omega}_i \bar{\sigma}_i$ over a given time window.
\end{itemize}
 These metrics are further described at the beginning of each corresponding sub-sections below.

\FloatBarrier
\section{Results}
\subsection{The effect of the bin size}

\begin{figure}
\centering
 \includegraphics[width=\linewidth]{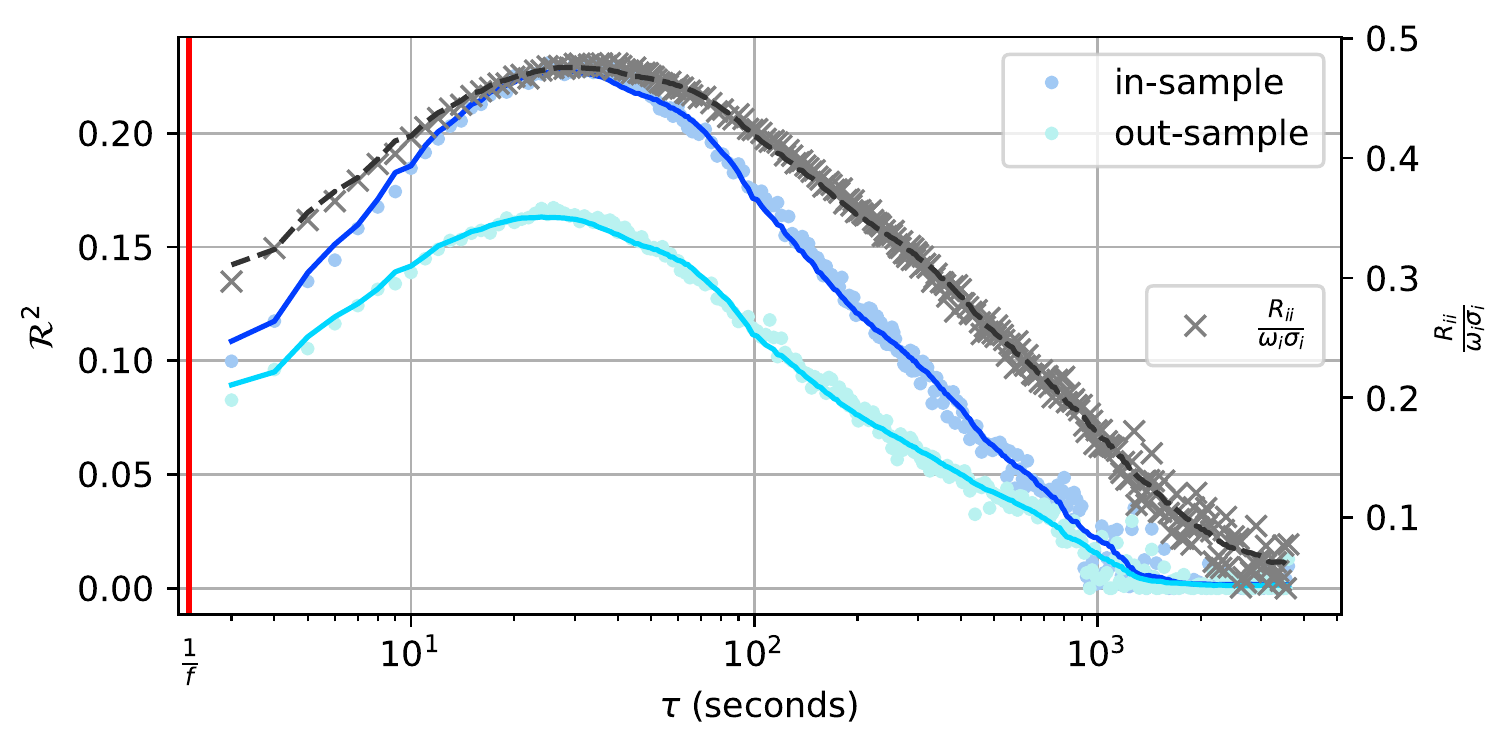}
 \caption{Single asset generalised R-squared~$\mathcal{R}^2(I_{\sigma_i})$ as a function of the bin size~$\tau$ for the asset~$TRMB$ (\textit{Trimble Navigation}). In-sample results were calibrated on the year $2021$, while out-of-sample outcomes cover $2022$. The vertical red line indicates the average time interval between two trades $1/f$.} 
 \label{fig:binning_test_R2_diag_first-stock}
\end{figure}

Within the linear framework previously defined, the market price impact of a single asset is a non-trivial function of the bin size~$\tau$ (Fig.~\ref{fig:binning_test_R2_diag_first-stock}). Specifically, the goodness-of-fit~$\mathcal{R}^2(I_{\sigma_i})$ increases with the bin size up to a maximum ranging generally between $10$ and $100$ seconds before decreasing down to a negligible level at $1$ hour. It is worth noting  that, for a single asset, the performances of the three previously mentioned models actually coincide. Indeed, these models all express price increments as a linear function of the signed order flow.

At short time scales, effects similar to the Epps effect \citep{Epps-1979,TothKertesz-2009} may prevent the correlation between the order flow and the price impact to become fully apparent without further corrections. Indeed, the correlation between the signed order flows and the price variations $\frac{R_{ii}}{\sigma_i \bar{\omega}_i}$ decreases when the time scale shortens (Fig.~\ref{fig:binning_test_R2_diag_first-stock}). In fact, this correlation is simply the square-root of the model accuracy~$\mathcal{R}^2(I_{\sigma_i})$. Yet, among the causes of the Epps effect (lead-lag effects, asynchronicity, and the minimal response time of traders) \citep{TothKertesz-2009}, only the third factor is deemed relevant. Indeed, for a given asset, its signed order flows and prices are updated synchronously with no lead-lag. More prosaically, when the bin size widens, the number of trades per bin increases and so does the accuracy of the model.

At larger time scales, the predictive power of the linear impact model decreases rapidly (Fig.~\ref{fig:binning_test_R2_diag_first-stock}). This decay cannot be attributed to the transient effect from the price impact (see section~\ref{Introduction}). Indeed, the magnitude of the price impact follows a power law with a slow decline, typically remaining significant after a thousand trades \citep{BouchaudEtAl-2018}. In the above example, the number of trades accumulated within one hour (the largest bin) is around $1400$. Consequently, it is reasonable to anticipate that the impact of the first trades within a bin would continue to be substantial at this time scale. However, we observe a decrease in the price impact model accuracy after a couple of minutes.

More generally, the impact from all the other trades, including the most recent ones, should be observable on the current price change in a bin, even at large time scales. However, as documented by \citet{PatzeltBouchaud-2017,PatzeltBouchaud-2018}, the relationship between the signed order flows and the price changes is actually reasonably fitted by a sigmoid function. This latter is indeed linear for reasonably small sizes of signed order flows. Yet, the non-linear relationship between price impacts and larger sizes of signed order flows, which is frequently observed in large-scale bins, explains the lower precision of linear models at these scales.

The relationship between the bin size~$\tau$ and the model accuracy $\mathcal{R}^2(M)$ provides an avenue to determine the maximum goodness-of-fit~$\mathcal{R}^{2*}(M)$ and its corresponding optimal time scale~$\tau^*(M)$. The ensuing sections investigate how these latter are influenced by the trading frequency of the assets, correlation among assets, and liquidity of the assets.

\FloatBarrier
\subsection{The effect of the trading frequency} \label{The effect of the trading frequency}

\subsubsection{Time scales} \label{The effect of the trading frequency: Time scales}

\begin{figure}
	\centering
	\includegraphics[width=\linewidth]{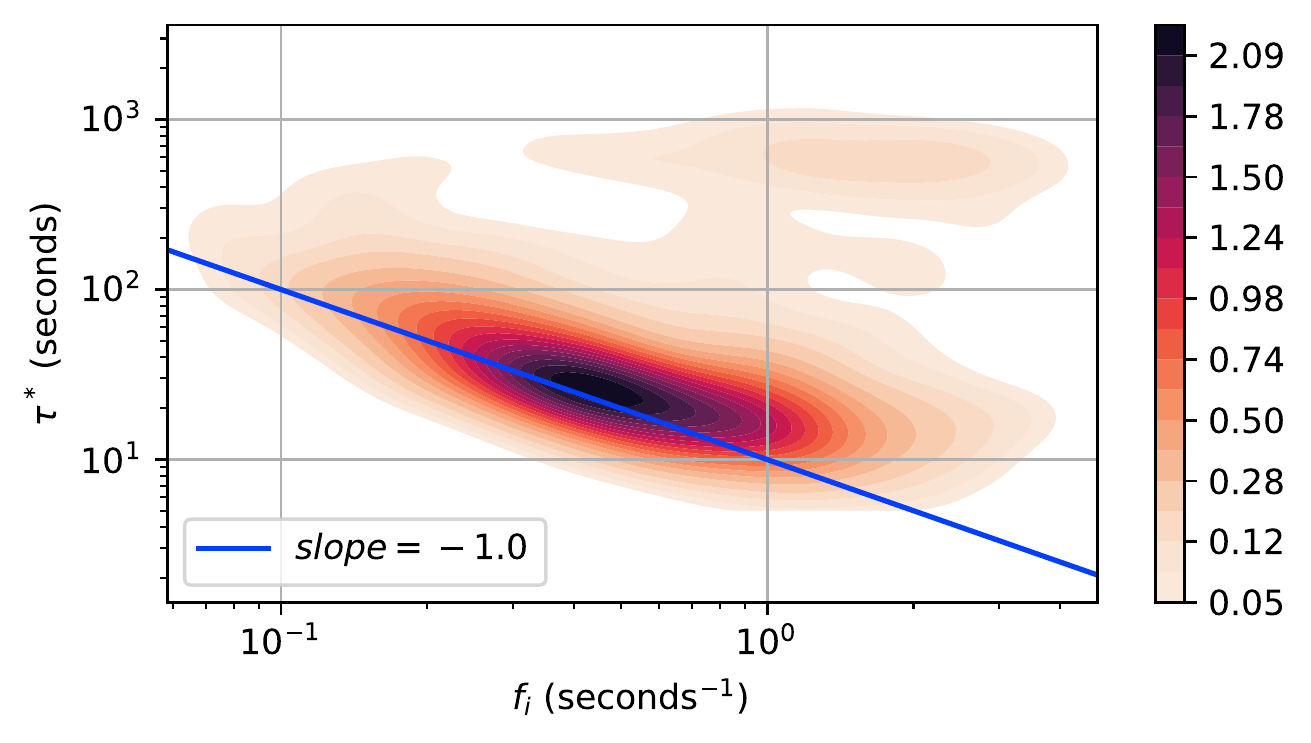}
	\caption{Empirical distribution of the optimal time scale out-of-sample~$\tau^*(I_{\sigma_i})$ as a function of the trading frequency~$f$ for single assets.}
	\label{fig:f/b_bin_size_R2_diag_first-stock_out_singles}
\end{figure}

For a given asset, we define its trading frequency~$f$ as its average number of trades per second during open market hours. Intuitively, one could expect higher trading frequencies to be associated with shorter optimal time scales~$\tau^*$, due to the quickest accumulation of trades in a bin. Empirically, higher trading frequencies do decrease the minimum optimal time scales achievable, yet other factors cause the considered assets to deviate from this limit. Specifically, the envelope of the normalized density plot associating optimal time scales to trading frequencies does not contain the lower left section of Fig.~\ref{fig:f/b_bin_size_R2_diag_first-stock_out_singles}. Thus, a minimum number of trades is required to reach the optimal bin size. Specifically, the blue straight line on this figure represents the function~$f\to 1/f$. Hence, one can estimate this minimum number of trades as the intercept ensuring that the majority of the data points are above this line. We find that a minimum of $10$ to $20$ trades is required to reach the optimal time scale of a linear cross-impact model.

The light brown area at the top right of Fig.~\ref{fig:f/b_bin_size_R2_diag_first-stock_out_singles} reveals a smaller group of assets with maximum cross-impact accuracy $\mathcal{R}^{2*}(I_{\sigma_i})$ for extended time scales. This group includes mainly large capitalization stocks (e.g. AMZN, AAPL, TSLA) and highly traded futures (E-mini S\&P futures, 10-year bond futures). For these assets, the causes decreasing the accuracy of the model seems to become effective for a larger number of trades ($500$ to $5000$ trades).

\FloatBarrier

\begin{figure} 
	\centering
	\subcaptionbox{As a function of $f_i$, the trading frequency of the predicted asset. \label{f_1/b_bin_size_delta_R2_kyle_first-stock_out_pairs} }{\includegraphics[width=0.49\linewidth]{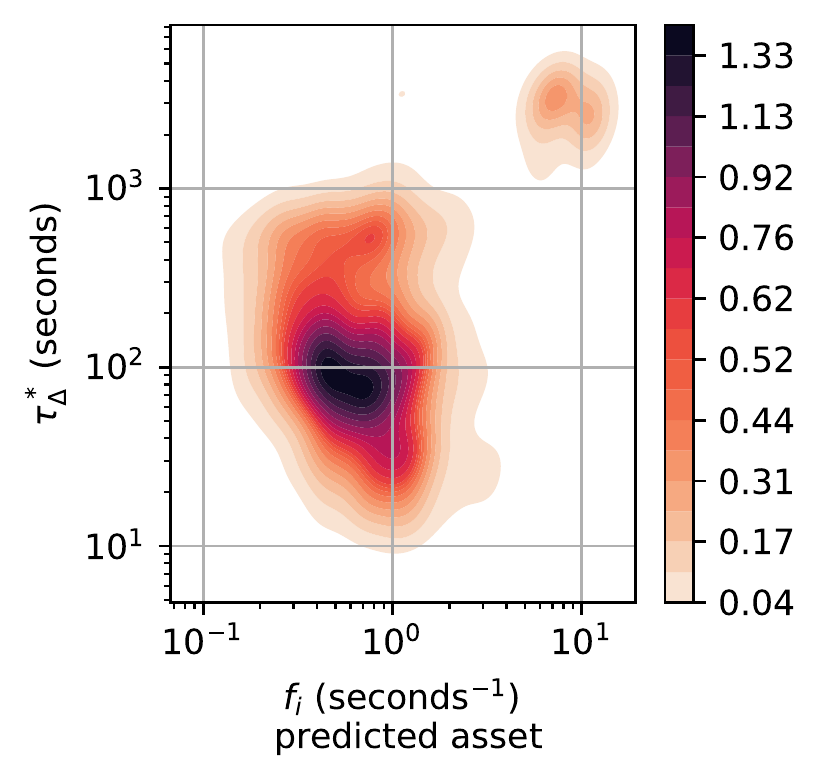}}
	\subcaptionbox{As a function of $f_j$, the trading frequency of the explanatory asset. \label{f_2/b_bin_size_delta_R2_kyle_first-stock_out_pairs}}{\includegraphics[width=0.49\linewidth]{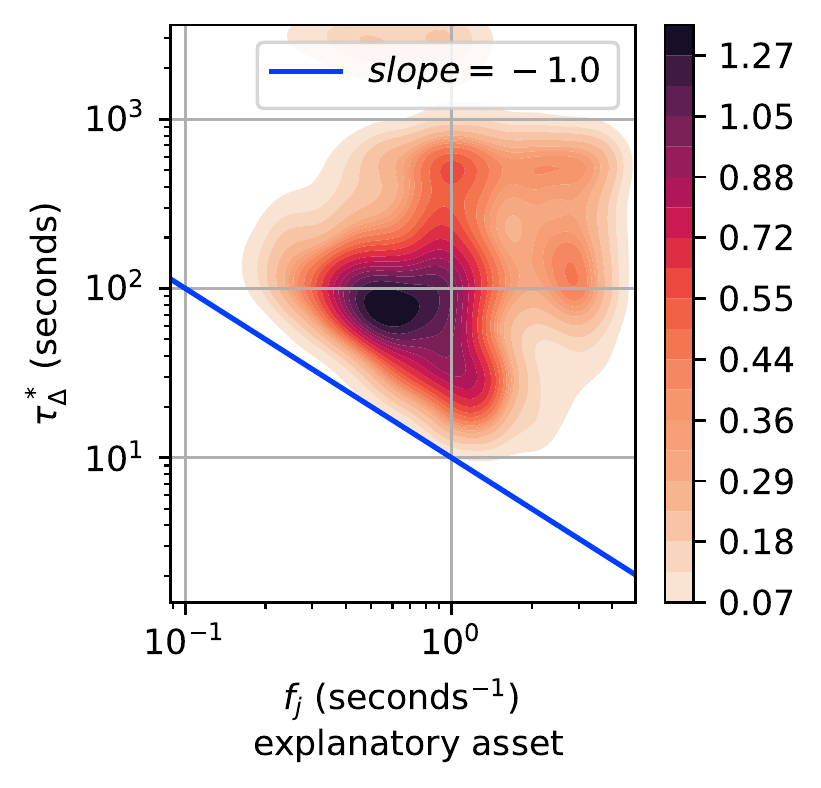}}
	\caption{Empirical distribution of the optimal time scale out-of-sample~$\tau^*_{\Delta}(I_{\sigma_i})$ as a function of the trading frequency of the predicted asset $f_i$ (Fig.~\ref{f_1/b_bin_size_delta_R2_kyle_first-stock_out_pairs}) or the explanatory asset $f_j$ (Fig.~\ref{f_2/b_bin_size_delta_R2_kyle_first-stock_out_pairs}). The asset pairs are filtered on the $7$\% of the sample exhibiting a correlation higher than $50$\% (see section~\ref{correlation}).}
	\label{fig:f_2/b_bin_size_delta_R2_kyle_first-stock_out_pairs}
\end{figure}

Figure~\ref{f_2/b_bin_size_delta_R2_kyle_first-stock_out_pairs} shows the impact of the trading frequency of the explanatory asset~$j$ regarding the goodness-of-fit on asset~$i$, when calibrating the model on pairs of assets. Specifically, $\tau^*_{\Delta}(I_{\sigma_i})$ corresponds to the time scale maximizing the added accuracy, $\Delta \mathcal{R}^{2*}(I_{\sigma_i})$, when predicting asset~$i$'s price increments. This indicator is driven by the trading frequency of the explanatory asset, as shown by the triangle form of Fig.~\ref{f_2/b_bin_size_delta_R2_kyle_first-stock_out_pairs}. As expected, this behavior cannot be observed when one increases the trading frequency of the predicted asset (Fig.~\ref{f_1/b_bin_size_delta_R2_kyle_first-stock_out_pairs}).

\begin{figure} 
	\centering
	\subcaptionbox{Mean deviation~$\left|\tau^*_{\Delta}(I_{\sigma_i})-\tau^*_{\text{diag}}(I_{\sigma_i})\right|$ by buckets of trading frequency gaps $\left|f_j-f_i\right|$. \label{log_abs_f_2-f_1/binning_peak_agg_abs_delta_tau__first-stock} }{\includegraphics[width=0.49\linewidth]{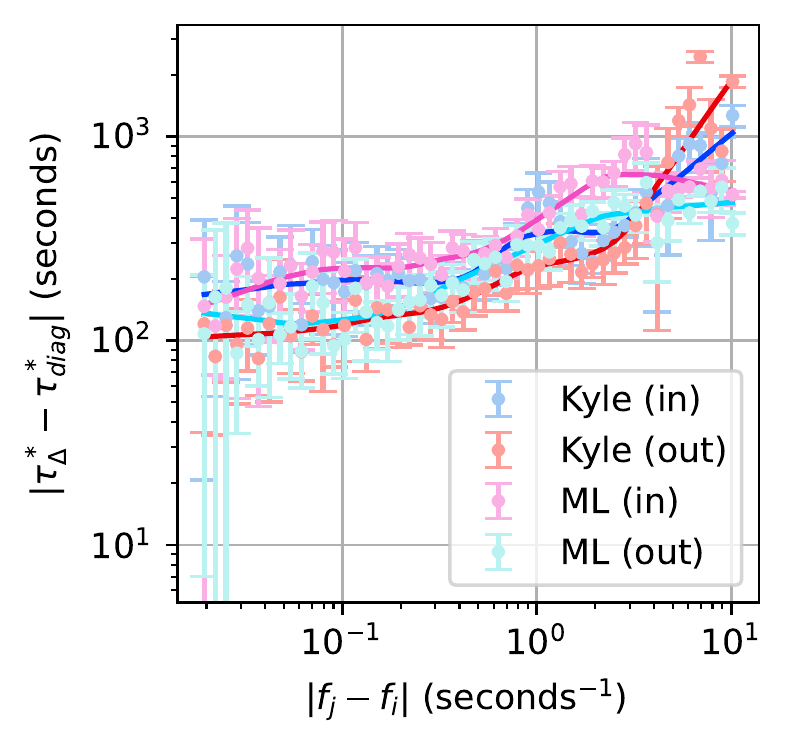}}
	\subcaptionbox{Empirical distribution of the deviation~$\left|\tau^*_{\Delta}(I_{\sigma_i})-\tau^*_{\text{diag}}(I_{\sigma_i})\right|$ in the Kyle model. \label{abs_f_2-f_1/abs_delta_tau__kyle_first-stock_out_pairs}}{\includegraphics[width=0.49\linewidth]{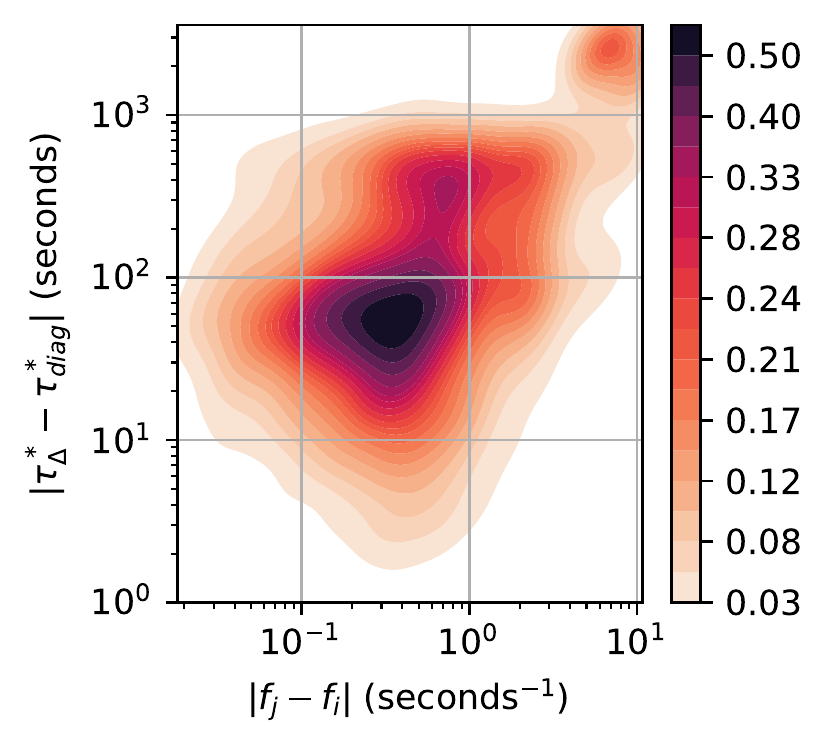}}
	\caption{Asynchronicity of the diagonal and cross sectional models $\left|\tau^*_{\Delta}(I_{\sigma_i})-\tau^*_{\text{diag}}(I_{\sigma_i})\right|$, aggregated by buckets of trading frequency gaps (Fig.~\ref{log_abs_f_2-f_1/binning_peak_agg_abs_delta_tau__first-stock}) or unaggregated (Fig.~\ref{abs_f_2-f_1/abs_delta_tau__kyle_first-stock_out_pairs}). The asset pairs are filtered on the $7$\% of the sample exhibiting a correlation higher than $50$\% (see section~\ref{correlation}).}
	\label{fig:log_abs_f_2-f_1/binning_peak_agg_abs_delta_tau__first-stock}
\end{figure}

On one hand, the optimal time scale for the cross-sectional effect is influenced by the trading frequency of the explanatory asset. On the other hand, the optimal time scale in the diagonal model is influenced by the frequency of the predicted asset. Consequently, significant deviations between the optimal time scales of the diagonal and cross section models are expected. As depicted in Fig.~\ref{fig:log_abs_f_2-f_1/binning_peak_agg_abs_delta_tau__first-stock}, these deviations increase when the trading frequencies of the two assets diverge. This may decrease the relevance of linear cross-impact models. Indeed, the optimal time scale of the cross sectional impact may be reached when the loss of accuracy from the direct price impact is higher than the marginal gain. However, this effect remains limited as demonstrated in Section~\ref{Multidimensional case}.

\begin{figure} 
	\centering
	\subcaptionbox{As a function of the maximum of the trading frequencies. \label{max_f_1_f_2/b_bin_size_delta_R2_kyle_first-stock_out_pairs} }{\includegraphics[width=0.49\linewidth]{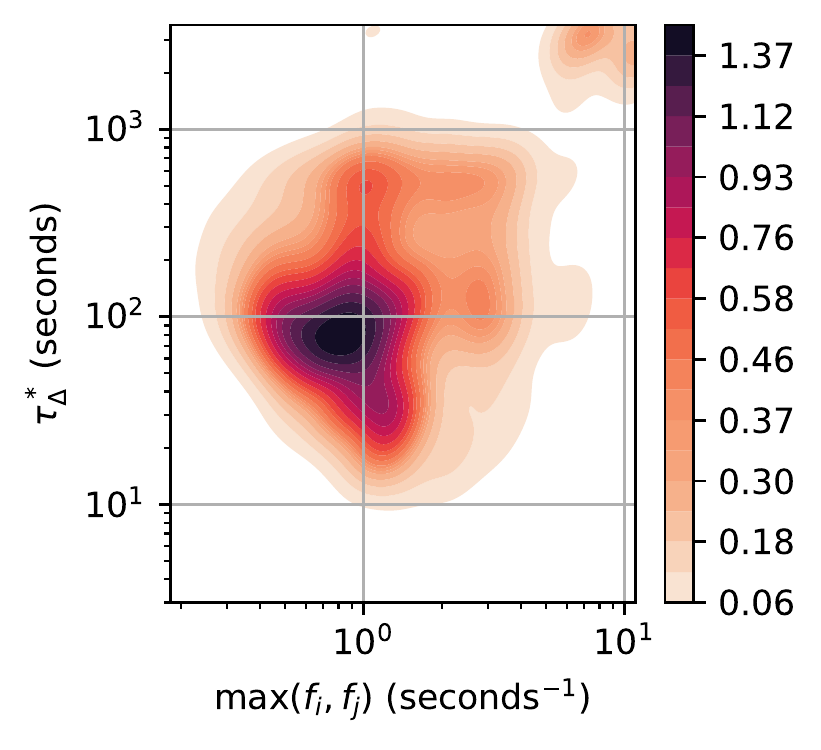}}
	\subcaptionbox{As a function of the minimum of the trading frequencies. \label{min_f_1_f_2/b_bin_size_delta_R2_kyle_first-stock_out_pairs}}{\includegraphics[width=0.49\linewidth]{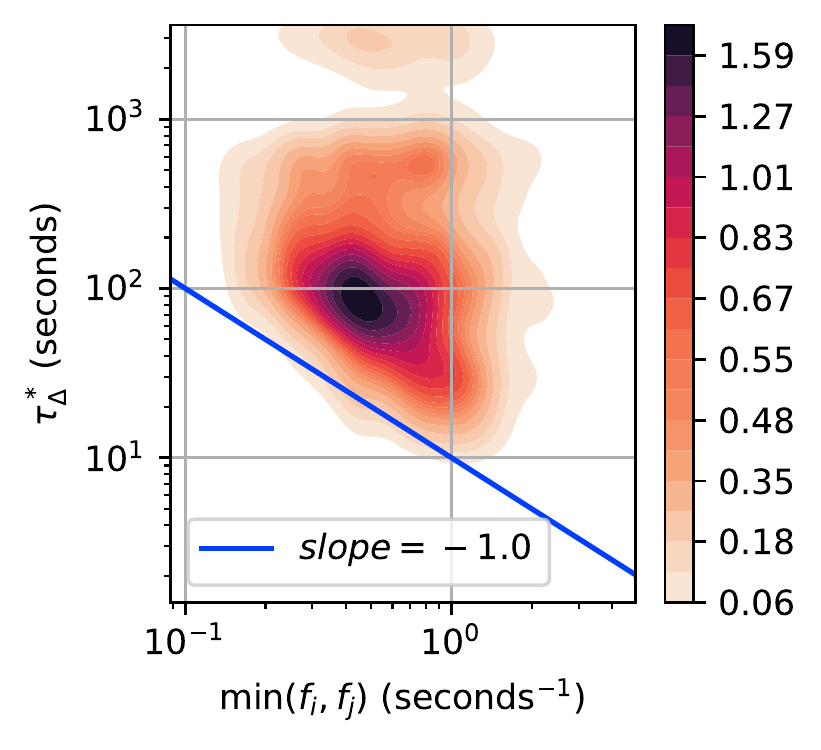}}
	\caption{Empirical distribution of the added accuracy optimal time scale out-of-sample~$\tau^*_{\Delta}(I_{\sigma_i})$ as a function of the maximum (Fig.~\ref{max_f_1_f_2/b_bin_size_delta_R2_kyle_first-stock_out_pairs}) or the minimum (Fig.~\ref{min_f_1_f_2/b_bin_size_delta_R2_kyle_first-stock_out_pairs}) of the trading frequencies of the assets pairs. The asset pairs are filtered on the $7$\% of the sample exhibiting a correlation higher than $50$\% (see section~\ref{correlation}).}
	\label{fig:max_f_1_f_2/b_bin_size_delta_R2_kyle_first-stock_out_pairs}
\end{figure}

\begin{figure} 
	\centering
	\subcaptionbox{As a function of the maximum of the trading frequencies. \label{max_f_1_f_2/b_bin_size_R2_kyle_first-stock_out_pairs} }{\includegraphics[width=0.49\linewidth]{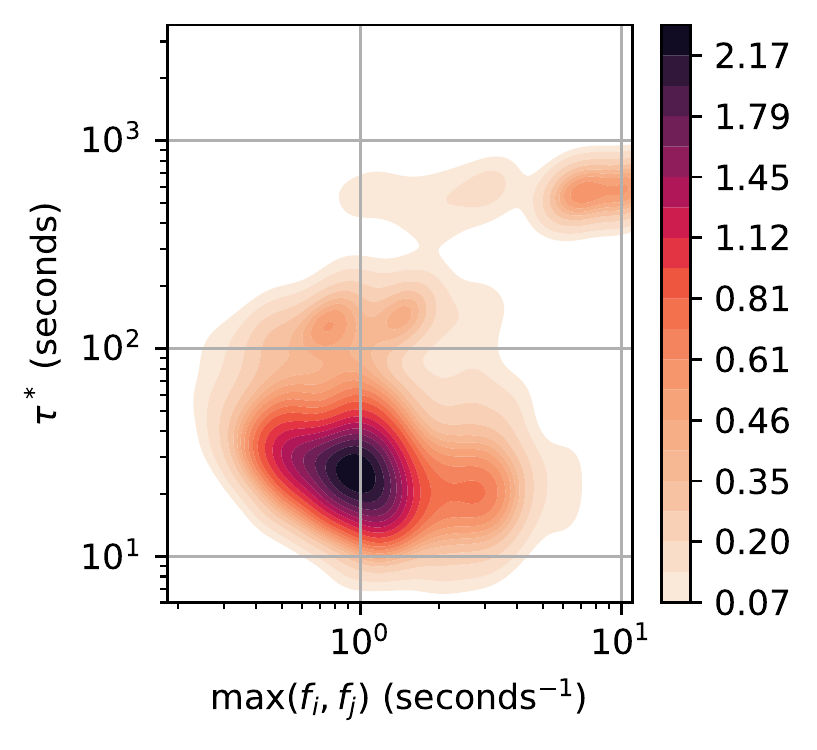}}
	\subcaptionbox{As a function of the minimum of the trading frequencies. \label{min_f_1_f_2/b_bin_size_R2_kyle_first-stock_out_pairs}}{\includegraphics[width=0.49\linewidth]{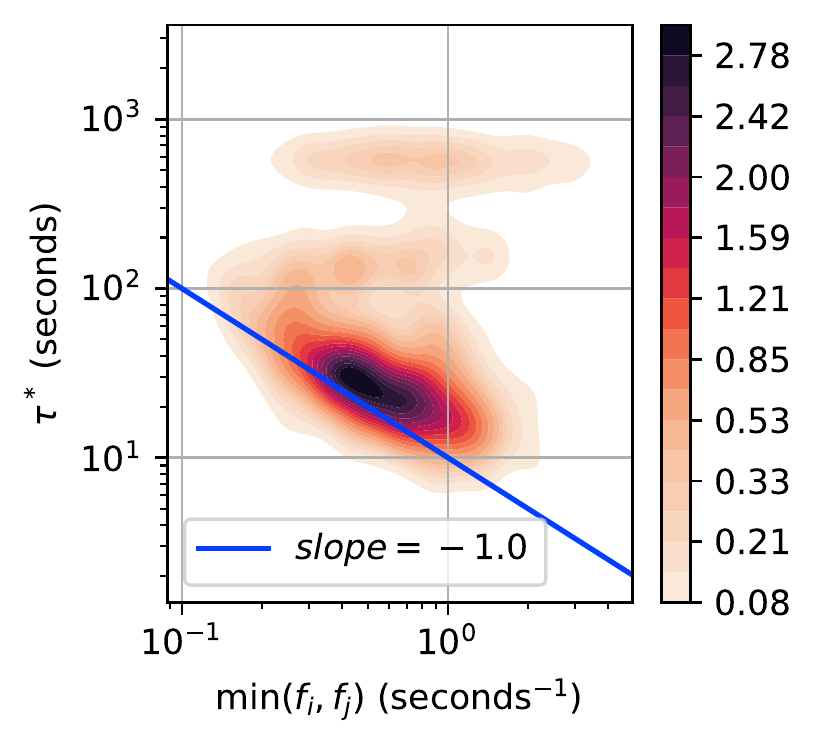}}
	\caption{Empirical distribution of the optimal time scale out-of-sample~$\tau^*(I_{\sigma_i})$ as a function of the maximum (Fig.~\ref{max_f_1_f_2/b_bin_size_R2_kyle_first-stock_out_pairs}) or the minimum (Fig.~\ref{min_f_1_f_2/b_bin_size_R2_kyle_first-stock_out_pairs}) of the trading frequencies of the assets pairs. The asset pairs are filtered on the $7$\% of the sample exhibiting a correlation higher than $50$\% (see section~\ref{correlation}).}
	\label{fig:max_f_1_f_2/b_bin_size_R2_kyle_first-stock_out_pairs}
\end{figure}

Finally, Fig.~\ref{min_f_1_f_2/b_bin_size_delta_R2_kyle_first-stock_out_pairs} demonstrates that the time scale~$\tau^*_{\Delta}(I_{\sigma_i})$ maximizing the added accuracy, is affected by the minimum of the trading frequencies of the assets pair. In contrast, the maximum of these two frequencies has little effect on this time scale (Fig.~\ref{max_f_1_f_2/b_bin_size_delta_R2_kyle_first-stock_out_pairs}). We observe the same behavior with respect to the optimal time scale~$\tau^*(I_{\sigma_i})$ (Fig.~\ref{fig:max_f_1_f_2/b_bin_size_R2_kyle_first-stock_out_pairs}). Consistently with the single asset case, we find that a minimum of $10$ to $20$ trades in both assets is required to reach the optimal time scale of a two-dimensional cross-impact model.

\FloatBarrier

\subsubsection{Goodness-of-fit}

\begin{figure} 
	\centering
	\subcaptionbox{Single assets in the diagonal model. \label{log_f-binning_peak_agg_m_R2_diag_first-stock} }{\includegraphics[width=0.49\linewidth]{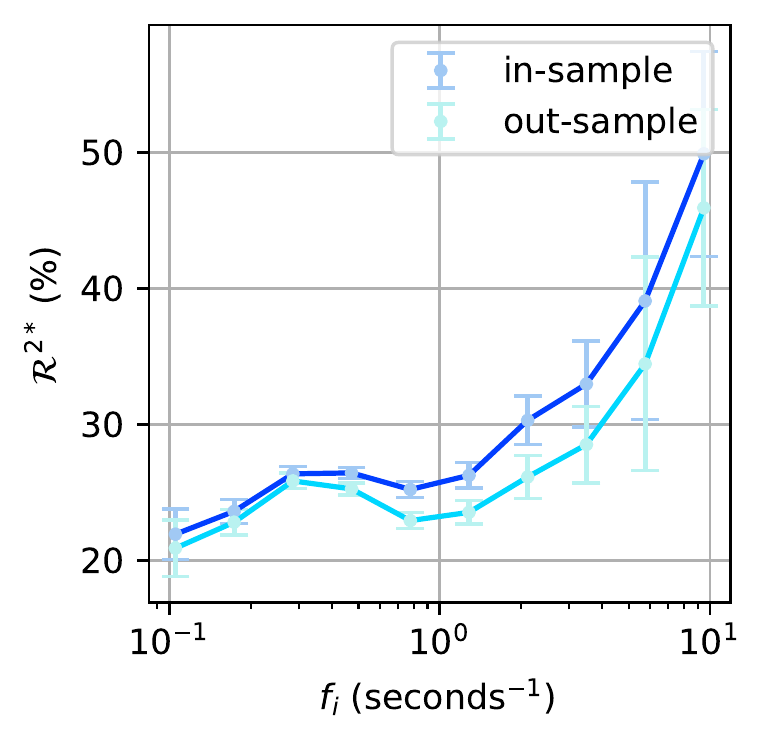}}
	\subcaptionbox{Pairs of assets. The asset pairs are filtered on the $7$\% of the sample exhibiting a correlation higher than $50$\%. \label{log_f_2/binning_peak_agg_m_delta_R2_mlt-stocks} }{\includegraphics[width=0.49\linewidth]{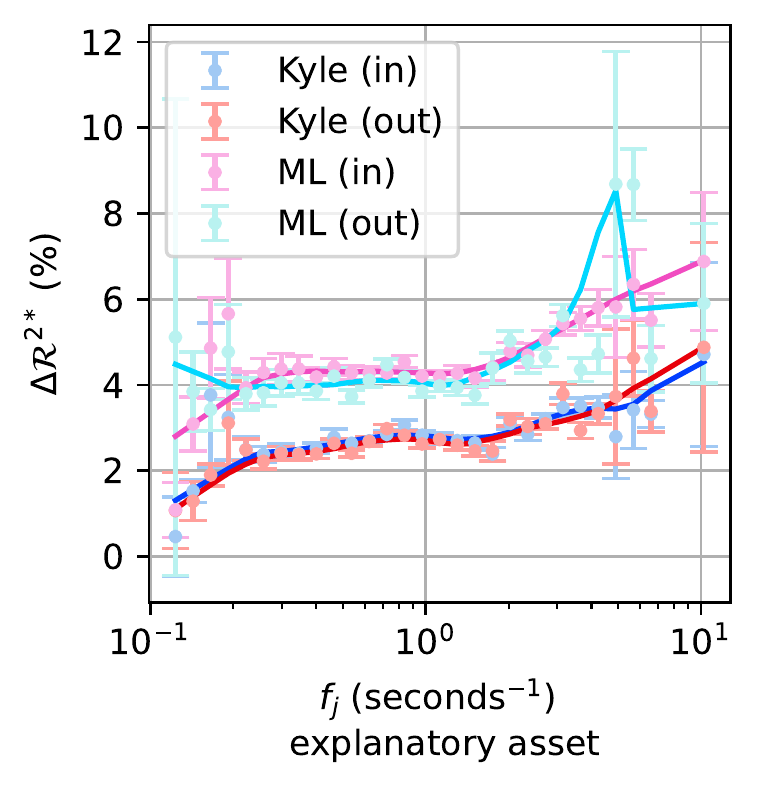}}
	\caption{Mean per trading frequency bucket of the optimal goodness-of-fit~$\mathcal{R}^{2*}(I_{\sigma_i})$ for single assets (Fig.~\ref{log_f-binning_peak_agg_m_R2_diag_first-stock}) and pairs of assets (Fig.~\ref{log_f_2/binning_peak_agg_m_delta_R2_mlt-stocks}).}
	\label{fig:log_f_2/binning_peak_agg_m_delta_R2_mlt-stocks}
\end{figure}

The trading frequency has a positive effect on the maximal accuracy~$\mathcal{R}^{2*}(I_{\sigma_i})$ observed across the tested bin sizes. Indeed, Fig.~\ref{fig:log_f_2/binning_peak_agg_m_delta_R2_mlt-stocks} shows that an increase in the trading frequency improves the mean~$\mathcal{R}^{2*}(I_{\sigma_i})$ per bucket. Here, the bars shaded in light pastel colors denote the range of two standard deviations surrounding the mean~$\mathcal{R}^{2*}(I_{\sigma_i})$ of the assets bucketed by trading frequency. The continuous lines in bright colors are the \textit{Locally Weighted Scatterplot Smoothing} \citep{Cleveland-1979} of these mean values. The following figures portraying bucketed data conform to the same convention.

The higher accuracy of the cross-impact model on highly traded assets can be attributed to the stronger correlation between prices and order flows when a sufficiently large number of market participants ensure the consistency of the two. However, it must be underlined that the impact of trading frequency is partially offset by observing the optimal accuracy~$\mathcal{R}^{2*}(I_{\sigma_i})$ across bin sizes, resulting in a relatively stable number of trades across trading frequencies. This effect probably explains the low slope in Fig.~\ref{fig:log_f_2/binning_peak_agg_m_delta_R2_mlt-stocks} when excluding data points with large error bars. 

Notice also the ML model out-of-sample performance in Fig.~\ref{log_f_2/binning_peak_agg_m_delta_R2_mlt-stocks} is significantly larger than that of the Kyle model, as previously reported in~\citet{TomasEtAl-2022}. The ML model can be easily over-fitted if one uses too little data, as it is more flexible than the Kyle model that imposes a no arbitrage condition. The fact that the out-of-sample performance of ML is better than Kyle shows that: (i) frictions in the market (bid-ask spread, fees) at least partially spoil the no-arbitrage assumption, as documented in~\citet{Schneider-2019}; (ii) this effect is significant enough to generalize well to yet unseen data.

\begin{figure} 
	\centering
	\subcaptionbox{Single assets in the diagonal model. \label{f/m_R2_diag_first-stock_out_singles}}{\includegraphics[width=0.49\linewidth]{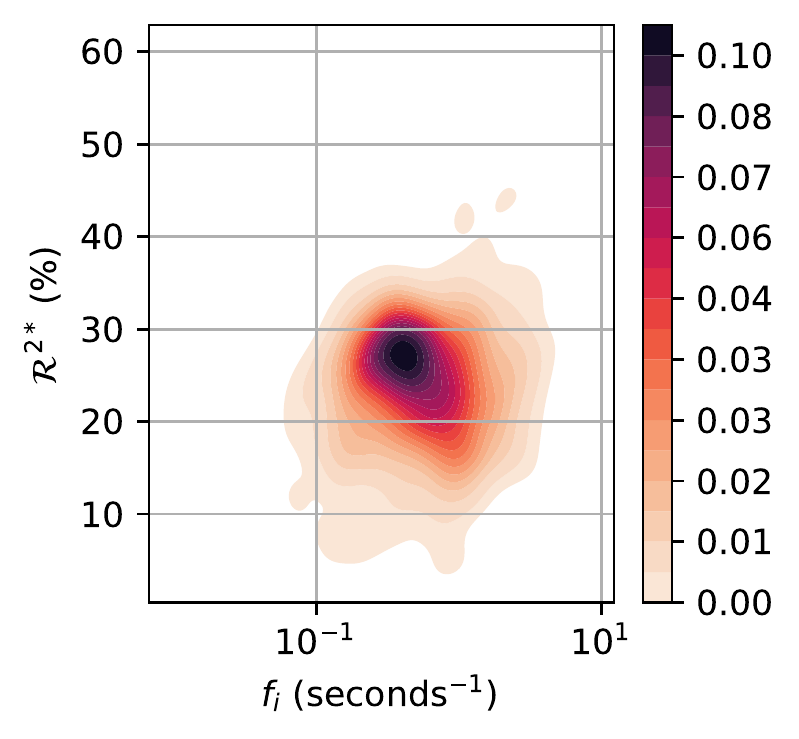}}
	\subcaptionbox{Pairs of assets in the Kyle model. The asset pairs are filtered on the $7$\% of the sample exhibiting a correlation higher than $50$\%.\label{f_2/m_delta_R2_kyle_first-stock_out_pairs} }{\includegraphics[width=0.49\linewidth]{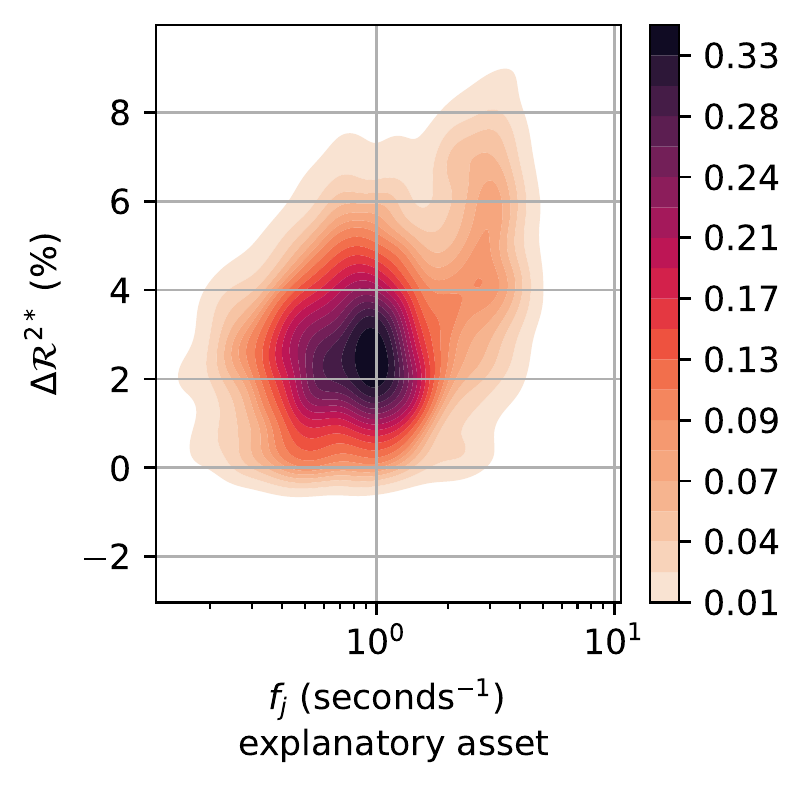}}
	\caption{Empirical distribution of the optimal out-of-sample~$\mathcal{R}^{2*}(I_{\sigma_i})$ as a function of the trading frequency~$f$ for single assets (Fig.~\ref{log_f-binning_peak_agg_m_R2_diag_first-stock}) and pairs of assets (Fig.~\ref{f_2/m_delta_R2_kyle_first-stock_out_pairs}).}
	\label{fig:f_2/m_delta_R2_kyle_first-stock_out_pairs}
\end{figure}

To provide a broader view of the distribution of~$\mathcal{R}^{2*}(I_{\sigma_i})$ across trading frequencies in our sample, we also present this data by density in Fig.~\ref{fig:f_2/m_delta_R2_kyle_first-stock_out_pairs}. The figure shows that most assets exhibit a trading frequency around $0.5$ trades per second, with an~$\mathcal{R}^{2*}(I_{\sigma_i})$ of $25\%$.

\FloatBarrier
\subsection{The effect of the correlation among assets} \label{correlation}

In this section, we examine the influence of asset correlations on cross-impact, emphasizing a clear distinction between these two concepts. This differentiation is pivotal since cross-impact effects reveal intricate relationships between correlation and liquidity. For example, when trading a diversified portfolio, execution costs are magnified on low liquidity factors regardless of correlation.

\begin{figure} 
\centering
 \includegraphics[width=\linewidth]{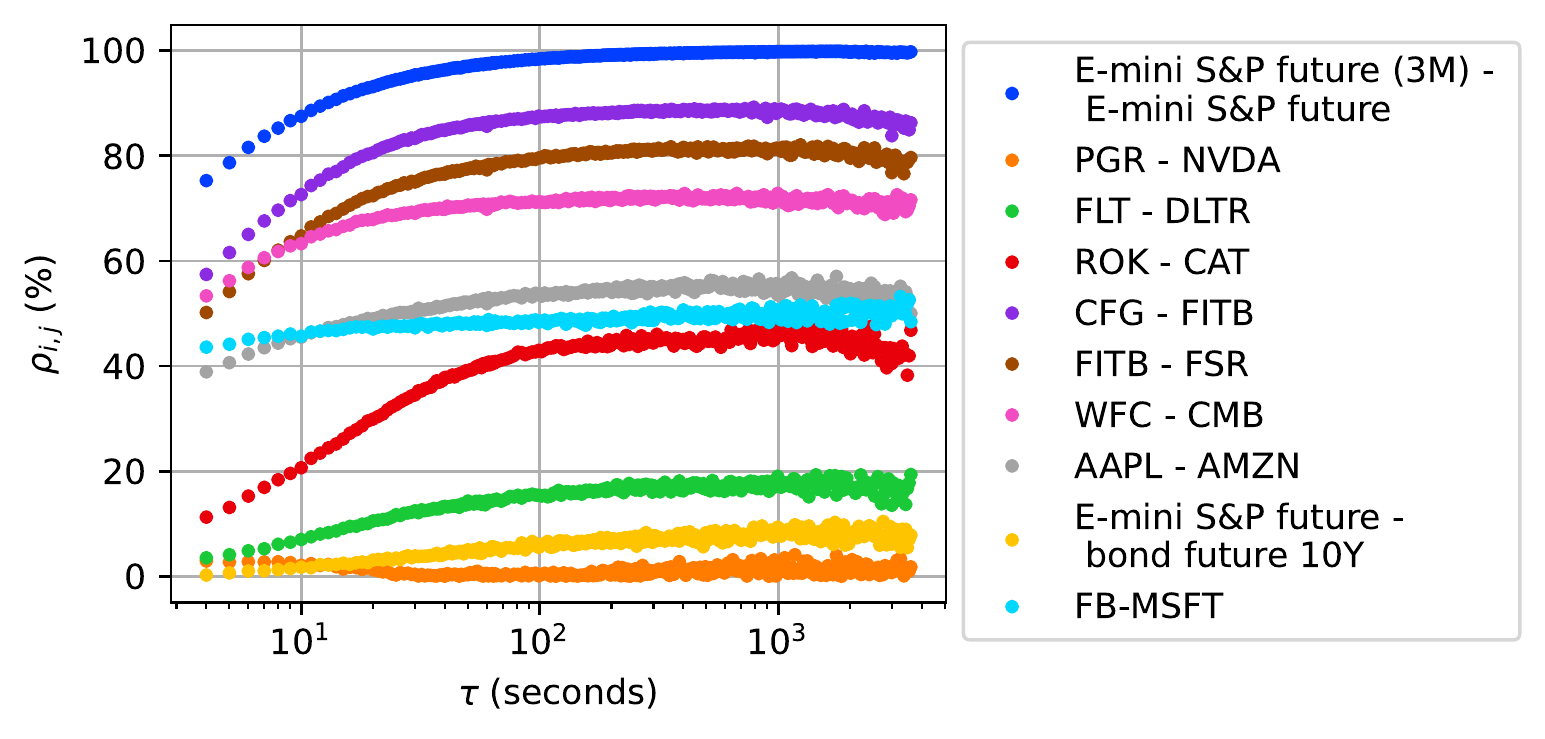}
 \caption{Pearson correlation coefficients~$\rho_{i,j}$ as a function of the bin size~$\tau$, for a selection of assets couples. Bin size can chosen arbitrarily small when utilizing  tick-by-tick data.} 
 \label{fig:df_epps_corr}
\end{figure}

Furthermore, it is important to note that correlations among assets are influenced by the time scale at which prices are sampled \citep{Epps-1979,Reno-2003,TothKertesz-2009}. Therefore, we use a bin size sufficiently large for the Epps effect to be negligible. Relying on the analysis presented on Fig.~\ref{fig:df_epps_corr}, we choose a $5$-minute bin. This time scale is a good compromise between Epps effect and noise.

\FloatBarrier
\subsubsection{Goodness-of-fit}

\begin{figure} 
	\centering
	\subcaptionbox{Empirical distribution of the out-of-sample $\Delta \mathcal{R}^{2*}(I_{\sigma})$ in the Kyle model. \label{corr/m_delta_R2_kyle_mlt-stocks_out_pairs}}{\includegraphics[width=0.49\linewidth]{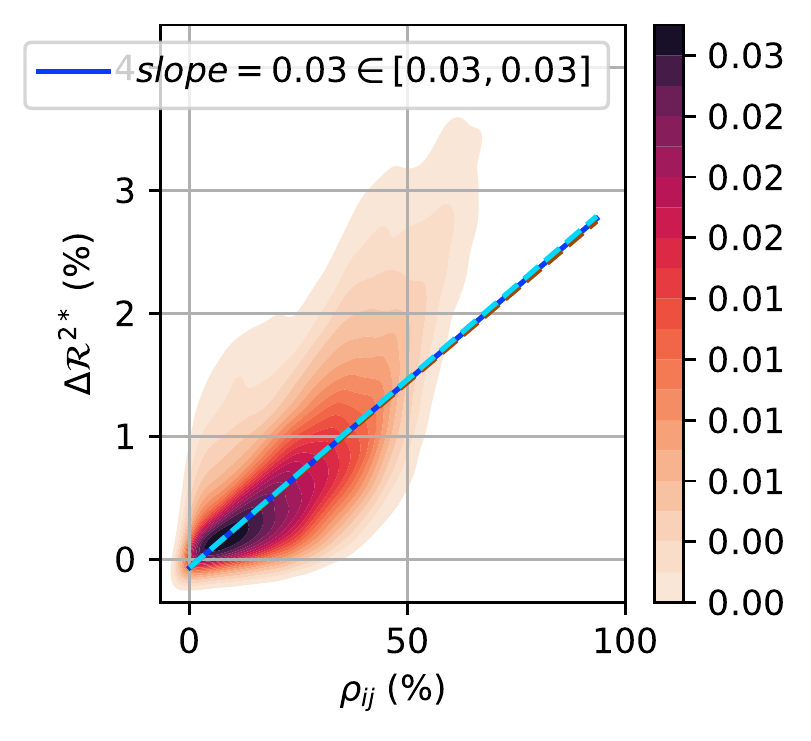}}
	\subcaptionbox{Mean per correlation bucket of the optimal $\Delta \mathcal{R}^{2*}(I_{\sigma})$. \label{corr/binning_peak_agg_m_delta_R2_mlt-stocks}}{\includegraphics[width=0.49\linewidth]{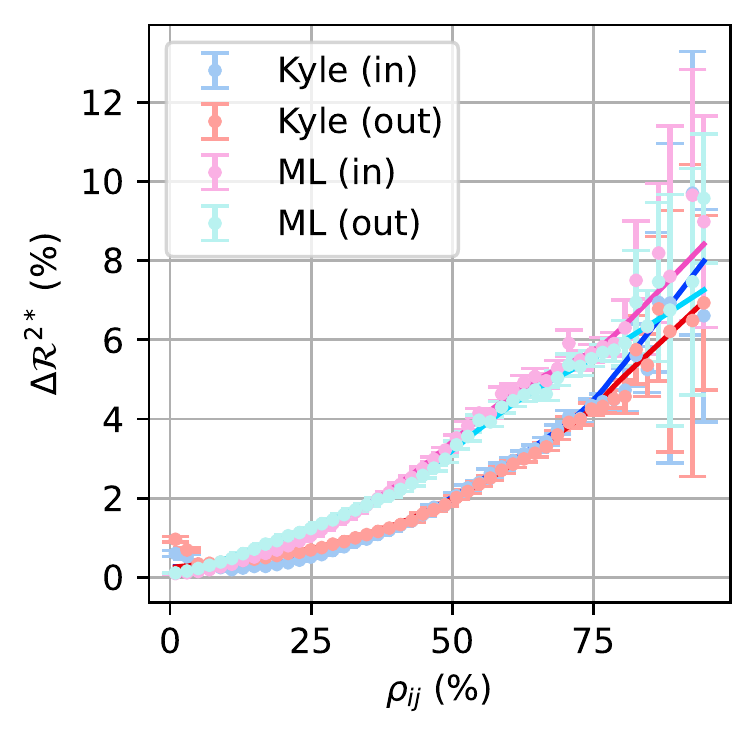}}
	\caption{Optimal \textbf{$\Delta \mathcal{R}^{2*}(I_{\sigma})$}, accounting for errors relative to both assets, as a function of the correlation~$\rho_{ij}$ for pairs of assets.}
	\label{fig:corr/binning_peak_agg_m_delta_R2_mlt-stocks}
\end{figure}

As expected, we observe a positive and monotonous relationship between the added accuracy from the cross sectional information $\Delta \mathcal{R}^{2*}(I_{\sigma})$ and the correlation~$\rho_{ij}$ among pairs of assets. For both the Kyle and ML models, $\Delta \mathcal{R}^{2*}(I_{\sigma})$ increases from 0 to above $5\%$ over the range of correlation levels in our sample (Fig.~\ref{fig:corr/binning_peak_agg_m_delta_R2_mlt-stocks}). Regarding Fig.~\ref{corr/m_delta_R2_kyle_mlt-stocks_out_pairs} and the following density plots, the continuous line slope represents the Theil-Sen estimator \citep{Theil-1992,Sen-1968,Siegel-1982}, while the dotted lines indicate the $95\%$ confidence interval around this estimate.

\subsubsection{Time scales}

\begin{figure} 
	\centering
	\subcaptionbox{Out-of-sample empirical distribution in the Kyle model. \label{corr/b_bin_size_R2_kyle_mlt-stocks_out_pairs}}{\includegraphics[width=0.49\linewidth]{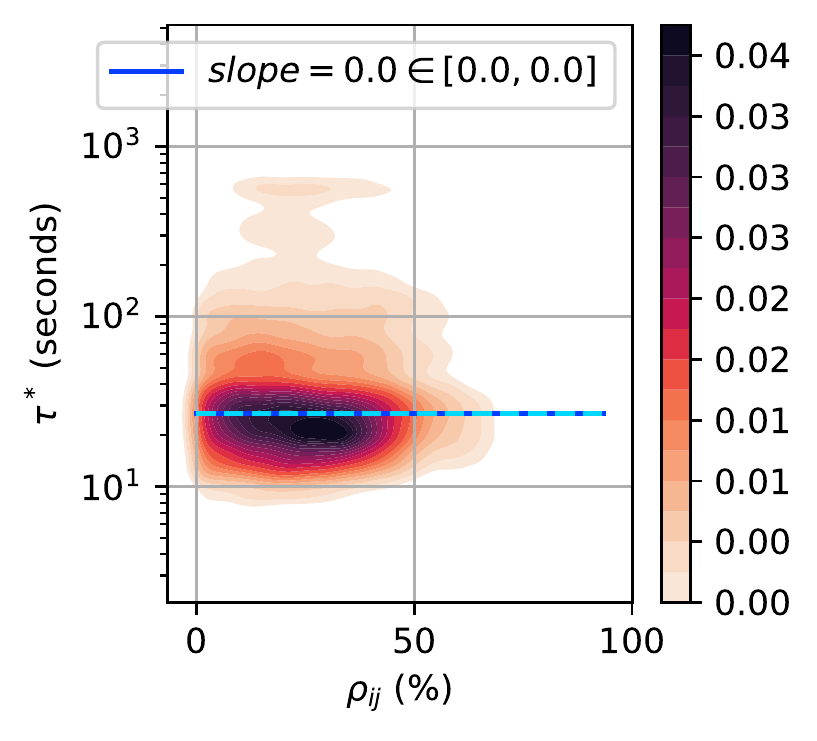}}
	\subcaptionbox{Mean per correlation bucket. \label{corr/binning_peak_agg_b_bin_size_R2_mlt-stocks}}{\includegraphics[width=0.49\linewidth]{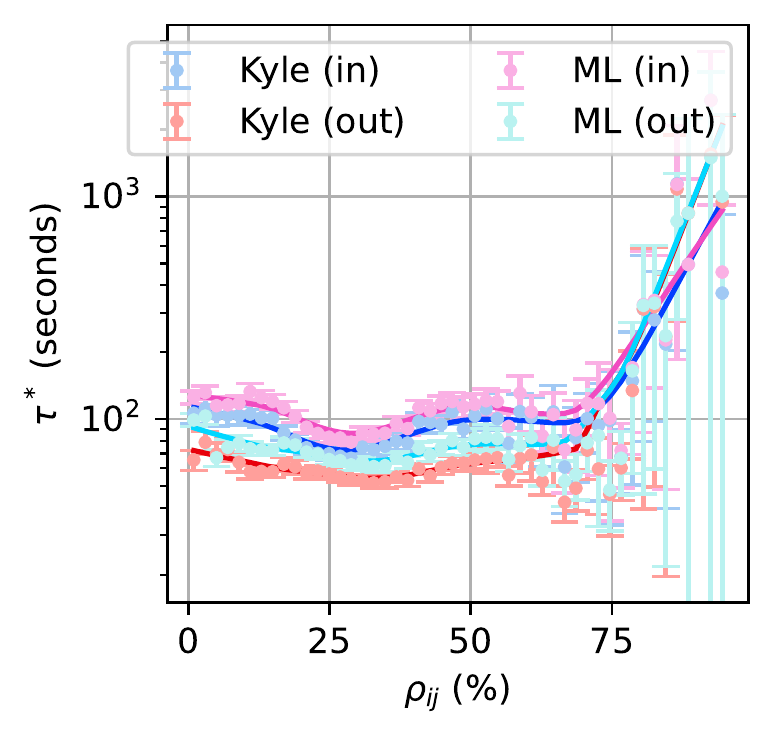}}
	\caption{Optimal time scale~$\tau^*(I_{\sigma})$ as a function of the correlation level~$\rho_{ij}$ for pairs of assets.}
	\label{fig:corr/binning_peak_agg_b_bin_size_R2_mlt-stocks}
\end{figure}

The optimal time scale~$\tau^*(I_{\sigma})$ seems unaffected by the correlation level~$\rho_{ij}$ among pairs of assets. Indeed, Fig.~\ref{corr/binning_peak_agg_b_bin_size_R2_mlt-stocks} shows that the mean value of $\tau^*_{\Delta}(I_{\sigma})$ is almost independent from the correlation level.

\FloatBarrier
\subsection{The effect of the liquidity} \label{The effect of liquidity}

\subsubsection{Goodness-of-fit}

The liquidity of each individual asset is measured using a risk indicator that represents the typical size of gains or losses during a given time interval. Specifically, the liquidity of the asset~$i$ is defined by $\bar{\omega}_i \bar{\sigma}_i$, estimated at a given bin size. We set the bin size to $5$ minutes, consistently with the binning frequency for the correlation estimation (see Section~\ref{correlation}).

\begin{figure} 
	\centering
	\subcaptionbox{Single assets. \label{log_risk_hmean-binning_peak_agg_m_R2_diag_first-stock} }{\includegraphics[width=0.49\linewidth]{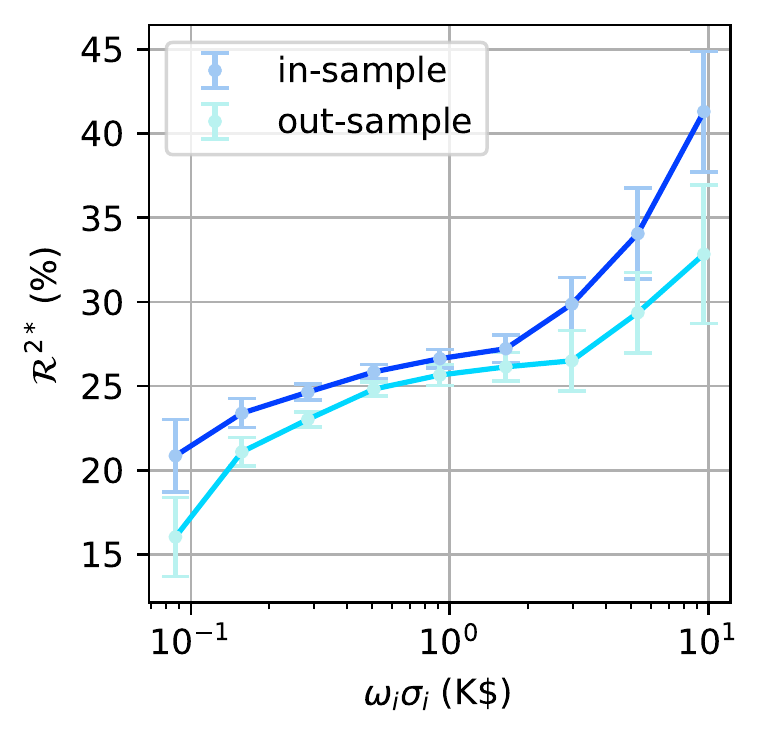}}
	\subcaptionbox{Pairs of assets filtered on the $7$\% of the sample exhibiting a correlation higher than $50$\%. \label{log_risk_2/binning_peak_agg_m_delta_R2_first-stock} }{\includegraphics[width=0.49\linewidth]{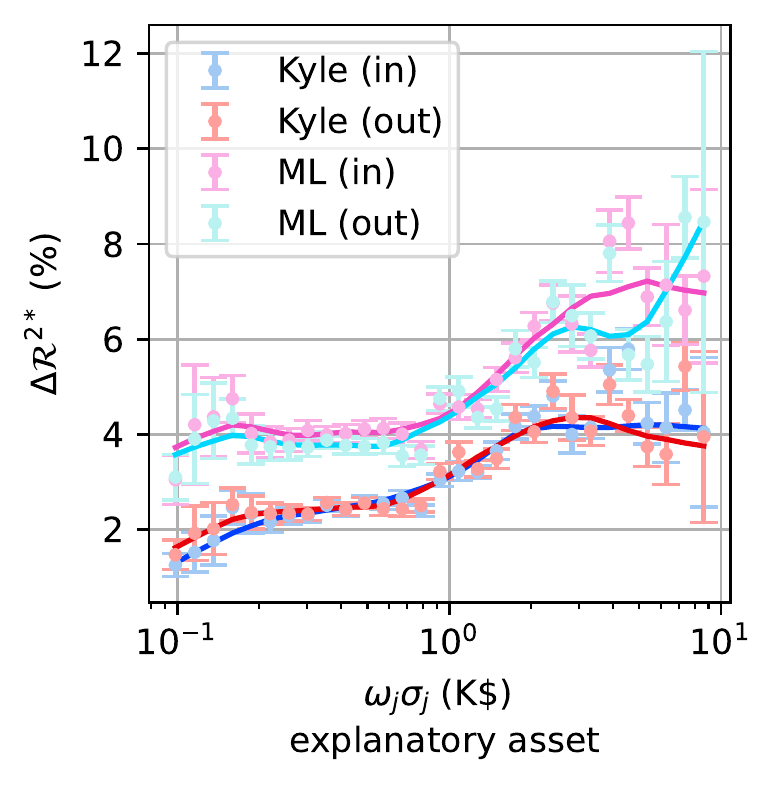}}
	\caption{Mean $\mathcal{R}^{2*}(I_{\sigma_i})$ by liquidity bucket for single assets (Fig.~\ref{log_risk_hmean-binning_peak_agg_m_R2_diag_first-stock}) and pairs of assets (Fig.~\ref{log_risk_2/binning_peak_agg_m_delta_R2_first-stock}).}
	\label{fig:log_risk_2/binning_peak_agg_m_delta_R2_first-stock}
\end{figure}

\begin{figure} 
	\centering
	\subcaptionbox{Single assets in the diagonal model. \label{risk-m_R2_diag_first-stock_out_singles} }{\includegraphics[width=0.49\linewidth]{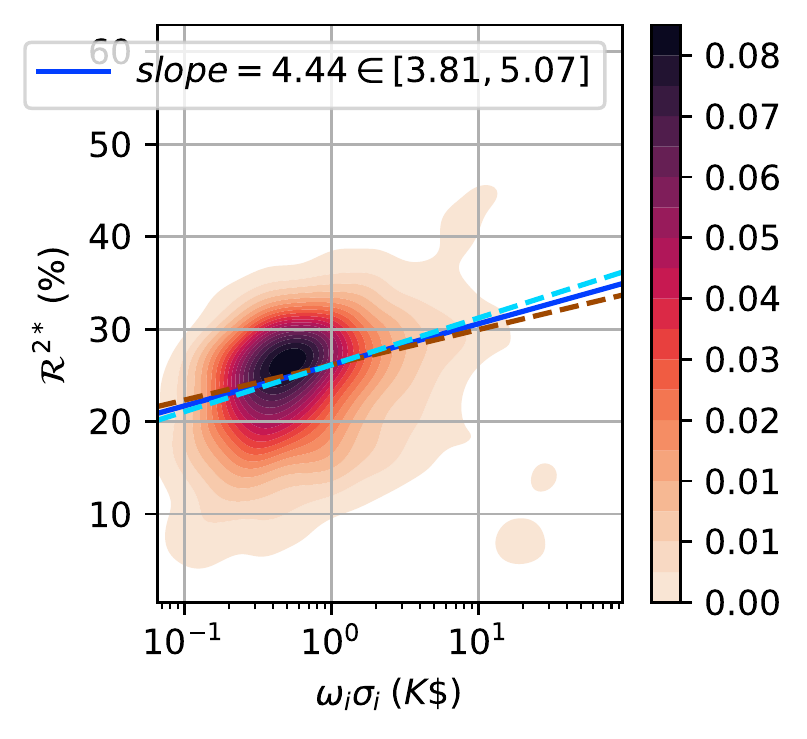}}
	\subcaptionbox{Pairs of assets filtered on the $7$\%\% of the sample exhibiting a correlation higher than $50$\%. \label{risk_2/m_delta_R2_kyle_first-stock_out_pairs} }{\includegraphics[width=0.49\linewidth]{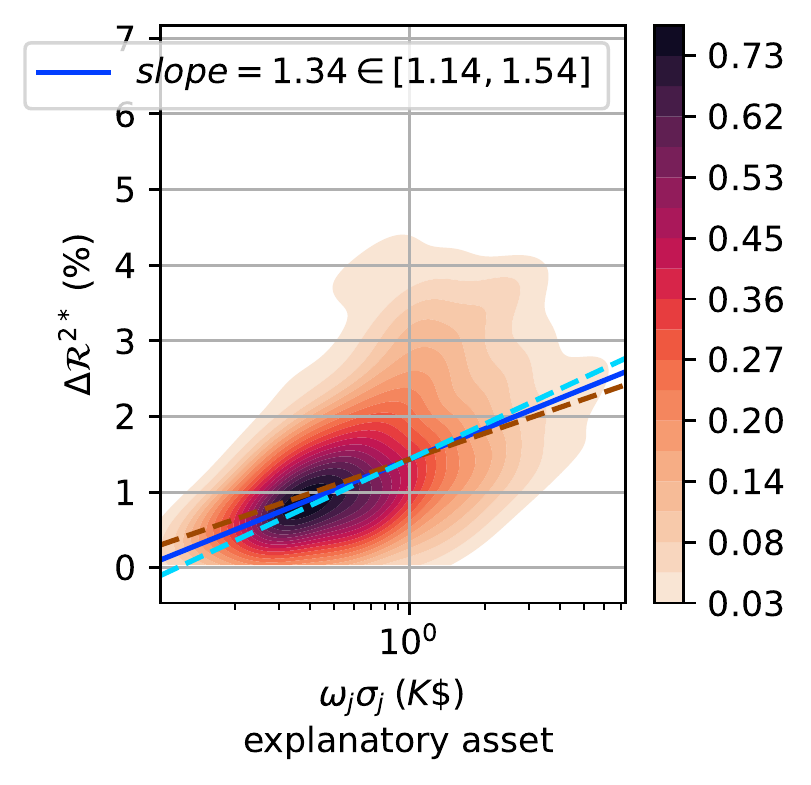}}
	\caption{Empirical distribution of the out-of-sample $\mathcal{R}^{2*}(I_{\sigma_i})$ as a function of the liquidity, for single assets (\ref{risk-m_R2_diag_first-stock_out_singles}) and pairs of assets (\ref{risk_2/m_delta_R2_kyle_first-stock_out_pairs}).}
	\label{fig:risk_2/m_delta_R2_kyle_first-stock_out_pairs}
\end{figure}

Liquidity has a positive effect on the accuracy of the cross-impact models tested, both for single assets and pairs of assets. Indeed, we observe that the out-of-sample goodness-of-fit increases from below $20$\% to around $30$\% across the liquidity levels in our sample for the diagonal model (Fig.~\ref{log_risk_hmean-binning_peak_agg_m_R2_diag_first-stock}). Like the interpretation proposed in Section~\ref{The effect of the trading frequency}, the higher score on liquid assets can be explained by the stronger correlation between prices and order flows when market liquidity is sufficient to ensure their consistency. As previously, we also present these results through density plots in Fig.~\ref{fig:risk_2/m_delta_R2_kyle_first-stock_out_pairs}. These plots illustrate that most assets exhibit a liquidity level of $500$ USD per $5$ minutes and an~$\mathcal{R}^{2*}(I_{\sigma_i})$ of around $25$\%.

\FloatBarrier

\subsubsection{Time scales} \label{The effect of liquidity: Time scales}

\begin{figure} 
	\centering
	\subcaptionbox{Single assets. \label{risk-b_bin_size_R2_diag_first-stock_out_singles} }{\includegraphics[width=0.46\linewidth]{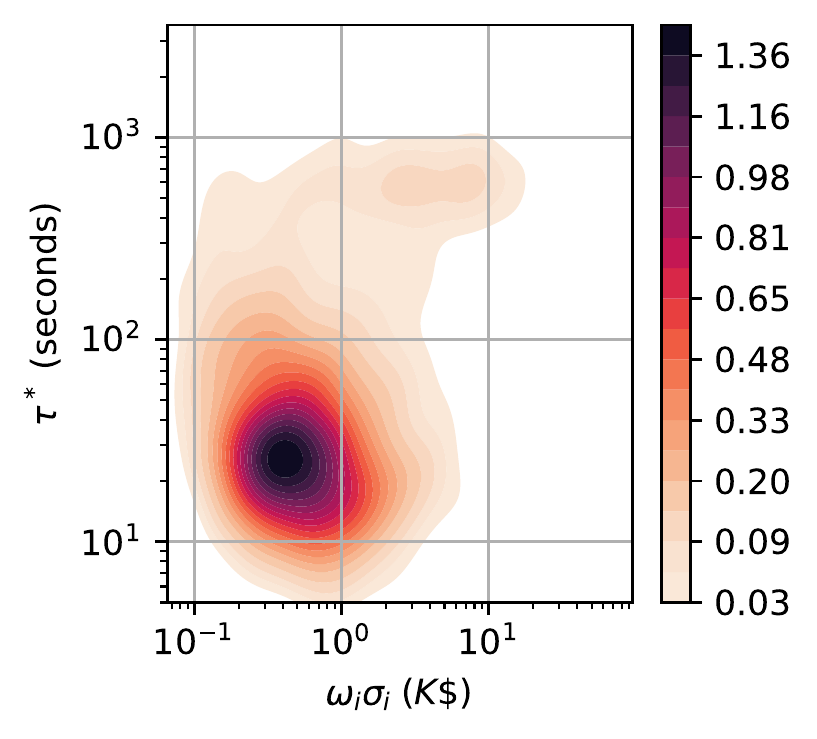}}
	\subcaptionbox{Pairs of assets in the Kyle model, filtered on the $7$\% of the sample exhibiting a correlation higher than $50$\%. \label{risk_2/b_bin_size_delta_R2_kyle_first-stock_out_pairs} }{\includegraphics[width=0.50\linewidth]{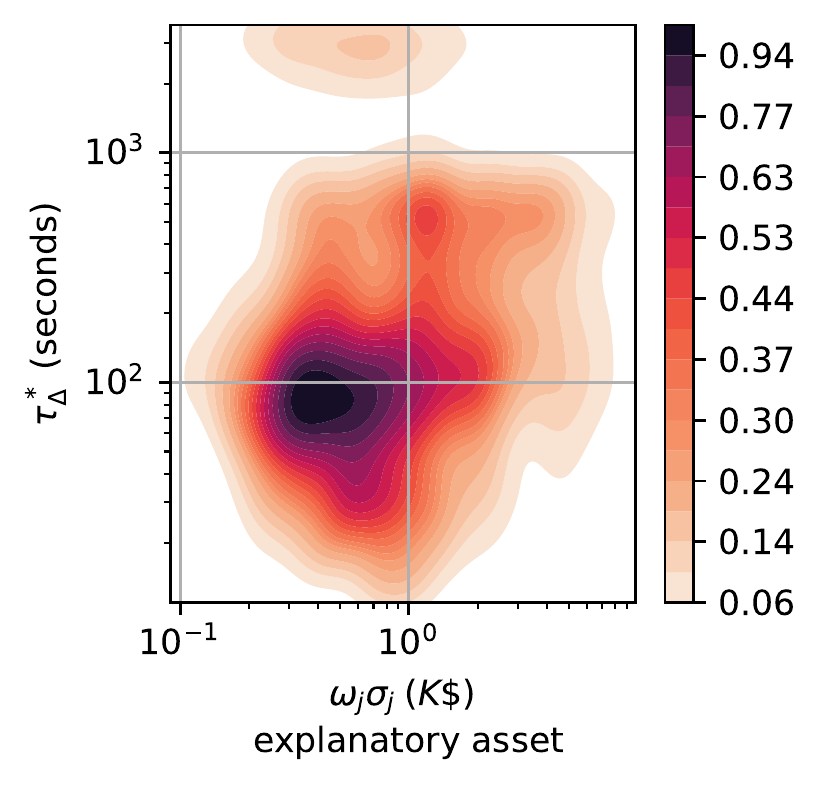}}
	\caption{Empirical distribution of the optimal bin size~$\tau^*(I_{\sigma_i})$ as a function of the liquidity, for single assets (Fig.~\ref{risk-b_bin_size_R2_diag_first-stock_out_singles}) and pairs of assets (Fig.~\ref{risk_2/b_bin_size_delta_R2_kyle_first-stock_out_pairs}).}
	\label{fig:risk_2/b_bin_size_delta_R2_kyle_first-stock_out_pairs}
\end{figure}

In contrast, liquidity has an ambiguous effect on the optimal time scale~$\tau^*(I_{\sigma_i})$. Notably, Fig.~\ref{risk_2/b_bin_size_delta_R2_kyle_first-stock_out_pairs} exhibits two groups of assets: (i) a large group with medium liquidity and time scales around $90$ seconds, (ii) a smaller group with higher liquidity and time scales around $10$ minutes. Within both groups, the liquidity seems to have a limited effect on the optimal time scale. These results suggest that other underlying properties of the considered assets influence the optimal time scale.

\FloatBarrier
\subsubsection{Cross effects of individual assets' liquidity}

\begin{figure} 
\centering
 \includegraphics[width=\linewidth]{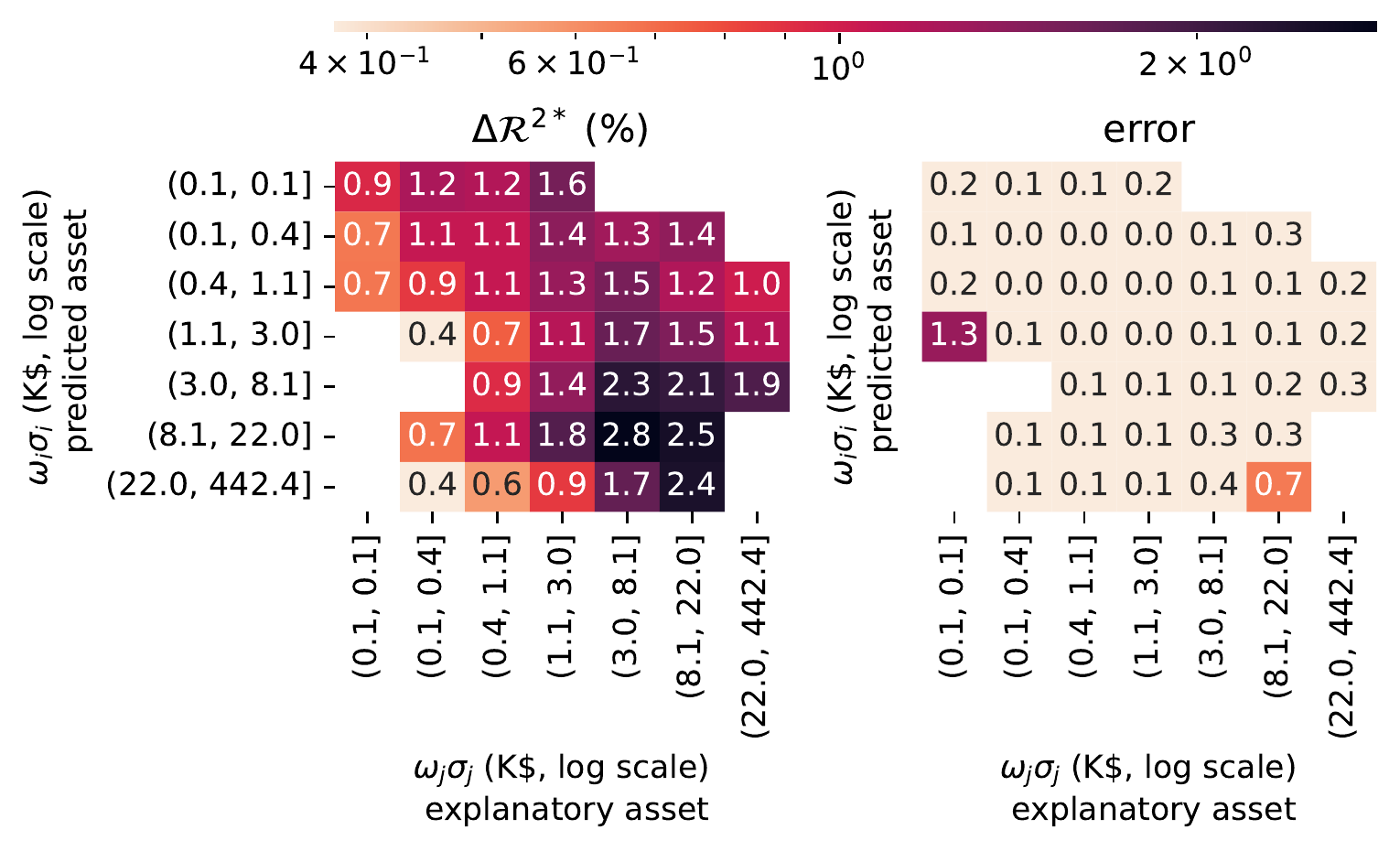}
 \caption{Mean out-of-sample added accuracy on asset~$i$ $\Delta \mathcal{R}^2(I_{\sigma_i})$, as a function of the individual risk levels of each asset in the Kyle model.}
 \label{fig:log_risk_1-log_risk_2-heatmap_m_delta_R2_kyle_first-stock_out}
\end{figure}

Our objective is to evaluate whether the added accuracy in the Kyle model is consistent with the stability properties outlined in Section~\ref{Models' properties}. Specifically, we seek to determine whether $\Delta \mathcal{R}^2(I_{\sigma_i})$ decreases as stock~$i$'s liquidity increases, but increases as stock~$j$'s liquidity increases, for a given pair~$(i,j)$ of assets. Notably, we expect that incorporating trades information from a high-liquidity asset to predict the prices of a low-liquidity asset will significantly enhance accuracy. However, this effect is not evident in Fig.~\ref{fig:log_risk_1-log_risk_2-heatmap_m_delta_R2_kyle_first-stock_out} due to the cross-effect of the correlation among assets. In fact, the greatest increase in accuracy for predicting asset~$i$'s prices is observed for relatively high levels of liquidity of asset~$i$. Empty cells in Fig.~\ref{fig:log_risk_1-log_risk_2-heatmap_m_delta_R2_kyle_first-stock_out} and the following heatmaps correspond to either an absence of assets in the associated buckets or to filtered-out results due to measurement errors being greater than than $50$\% of the mean value. As previously, these errors are defined by two standard deviations.

\begin{figure} 
\centering
\includegraphics[width=\linewidth]{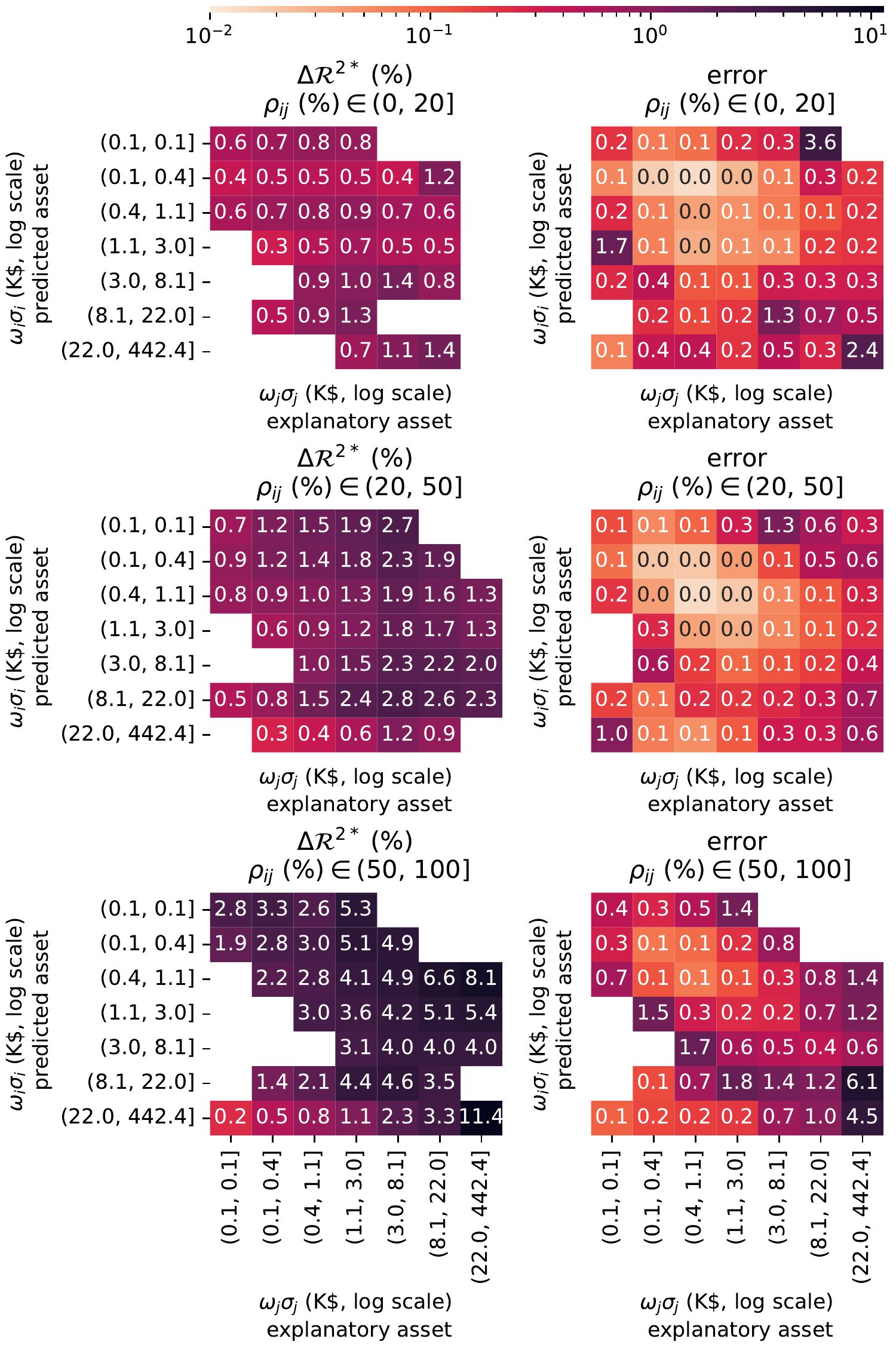}
\caption{Mean out-of-sample added accuracy on asset~$i$ $\Delta \mathcal{R}^2(I_{\sigma_i})$, as a function of the individual risk levels of each asset in the Kyle model.}
\label{fig:mlt_heatmap}
\end{figure}

Thus, we neutralize the effect of the correlation in Fig.~\ref{fig:mlt_heatmap} by grouping the asset pairs of our sample into correlation buckets. While at low correlations levels (from $0\%$ to $20$\%) the lowest liquid asset remains poorly predicted by the highest liquid asset, the effect of the cross sectional information becomes significant when looking at pairs of well correlated assets (above $50\%$). In a nutshell, cross-impact is significant if the predicted asset has a lower liquidity than the explanatory asset.

\subsection{Discussion}
Cross-impact is not relevant under all circumstances. Following this analysis, one can establish three requirements to accurately predict prices from cross sectional trades. Firstly, cross-impact does not occur at every time scale. A minimum number of trades in both assets, between $10$ to $20$, is required to observe a significant added accuracy on the prediction of asset prices. Secondly, trades only significantly explain the prices of highly correlated assets (correlations higher than $50$\%), regardless of the time scale. Thirdly, cross-impact explains a larger share of price variances if the predicted asset has a lower liquidity than the explanatory asset.

The previous section also establishes that on a pairwise basis, cross-impact is not a dominant effect among instruments with small correlations and comparable liquidity. This is in line with the results of \citet{ContEtAl-2023}, showing that on US stock markets, where correlations are on average quite low (i.e. 25 $\%$ in our sample for the $2017-2022$ period), pairwise cross-impact does not explain a large part of the price variance. However, cross-impact being small on a pairwise basis does not imply it remains a subdominant effect along factors or portfolio, as reported in \citet{TomasEtAl-2022} and \citet{BenzaquenEtAl-2017}. Indeed, the aggregation of flow from a large number of weakly correlated instruments can still lead to a significant increase of explanatory power for the price of a factor or a complex portfolio \citep{ContEtAl-2023}. The next section will illustrate this effect in the case of bonds.

To conclude, we can draw the following narrative. Price formation occurs endogenously within highly liquid assets. Then, trades in these assets influence the prices of their less liquid correlated products, with an impact speed constrained by their minimum trading frequency.

\FloatBarrier
\section{Application to the interest rate curve}

\subsection{Assets pairs}

\begin{figure} 
\centering
\includegraphics[width=\linewidth]{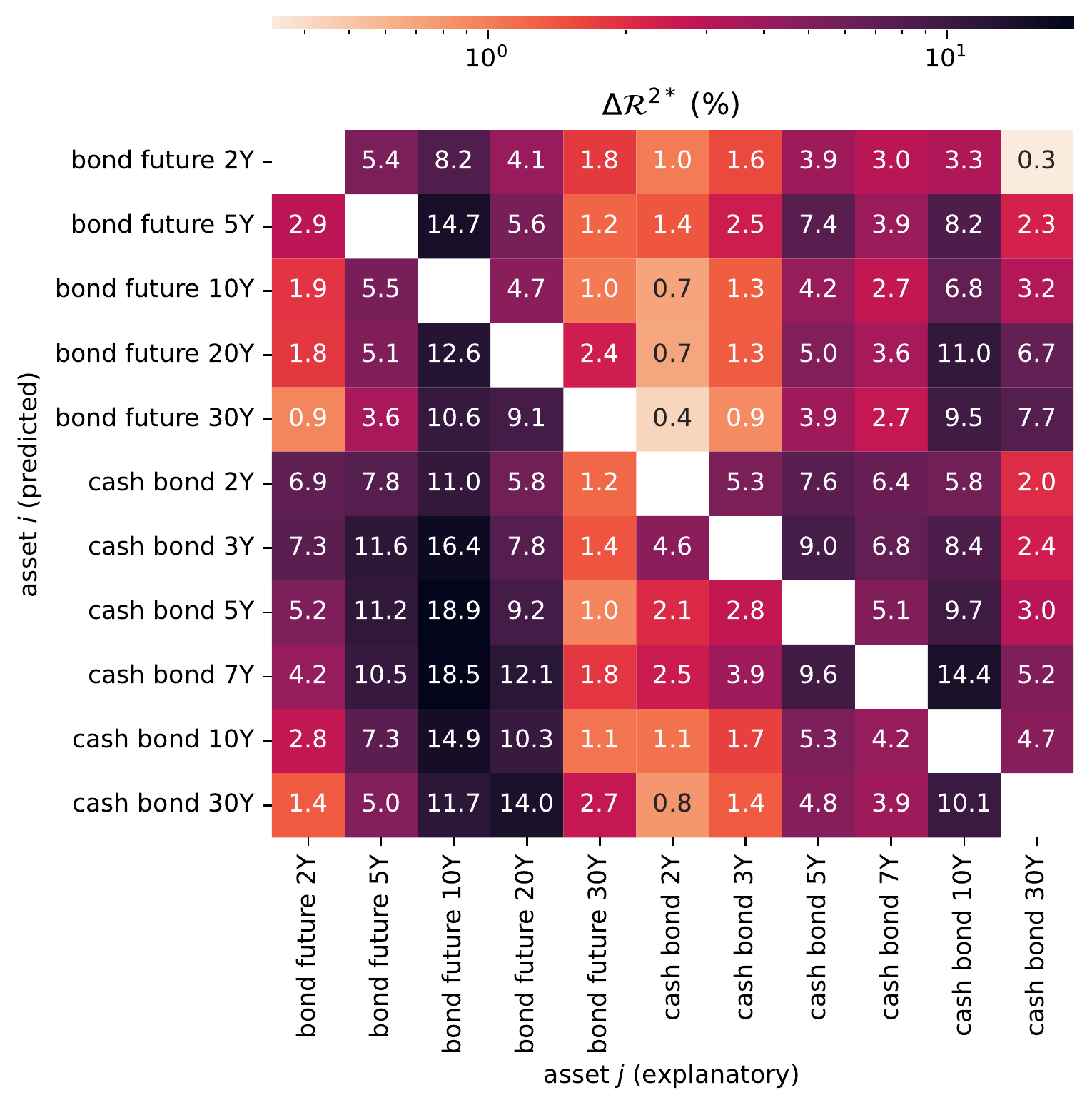}
\caption{Out-of-sample added accuracy $\Delta \mathcal{R}^2(I_{\sigma_i})$ in the Kyle model for each pair of assets of the interest rate curve.}
\label{fig:product_1-product_2/heatmap_m_delta_R2_kyle_first-stock_out}
\end{figure}

According to the previously established narrative, the interest rate curve should be a good candidate to apply a cross-impact model. Indeed, bonds of different tenors are highly correlated and display a wide range of liquidity levels. In this context, we run our previous analysis on a restriction of our initial sample to sovereign cash bonds and bonds futures in the United States for the $2021-2022$ period. Figure~\ref{fig:product_1-product_2/heatmap_m_delta_R2_kyle_first-stock_out} shows that including the trades of the most liquid assets (the 10-year future and cash bond) significantly increases the prediction accuracy concerning most of the other less liquid tenors.

Of particular significance, Fig.~\ref{fig:product_1-product_2/heatmap_m_delta_R2_kyle_first-stock_out} reveals that the trading information transmission flows from the most liquid tenors to that of lower liquid. This behavior challenges the validity of the theory in Financial Economics that regards long-term rates as agents anticipations of future short term rates. In practice, the prices of the low-liquidity tenors are more strongly impacted by the trades of the high-liquidity tenors than vice-versa (e.g. the 2-year cash bond in Fig.~\ref{fig:product_1-product_k/heatmap__R2_kyle_first-stock_out}). Future work could be devoted to the extension of this analysis to repurchase agreements of shorter tenors (such as 1-day, 1-week and 1-month tenors).

Because of the correlation among assets, the total added accuracy from using all asset trades information is not the summation of a given row of the matrix displayed in Fig.~\ref{fig:product_1-product_2/heatmap_m_delta_R2_kyle_first-stock_out}. Therefore, we display the multidimensional case in the next section.

\subsection{Multidimensional case} \label{Multidimensional case}

\begin{figure} 
\centering
\includegraphics[width=\linewidth]{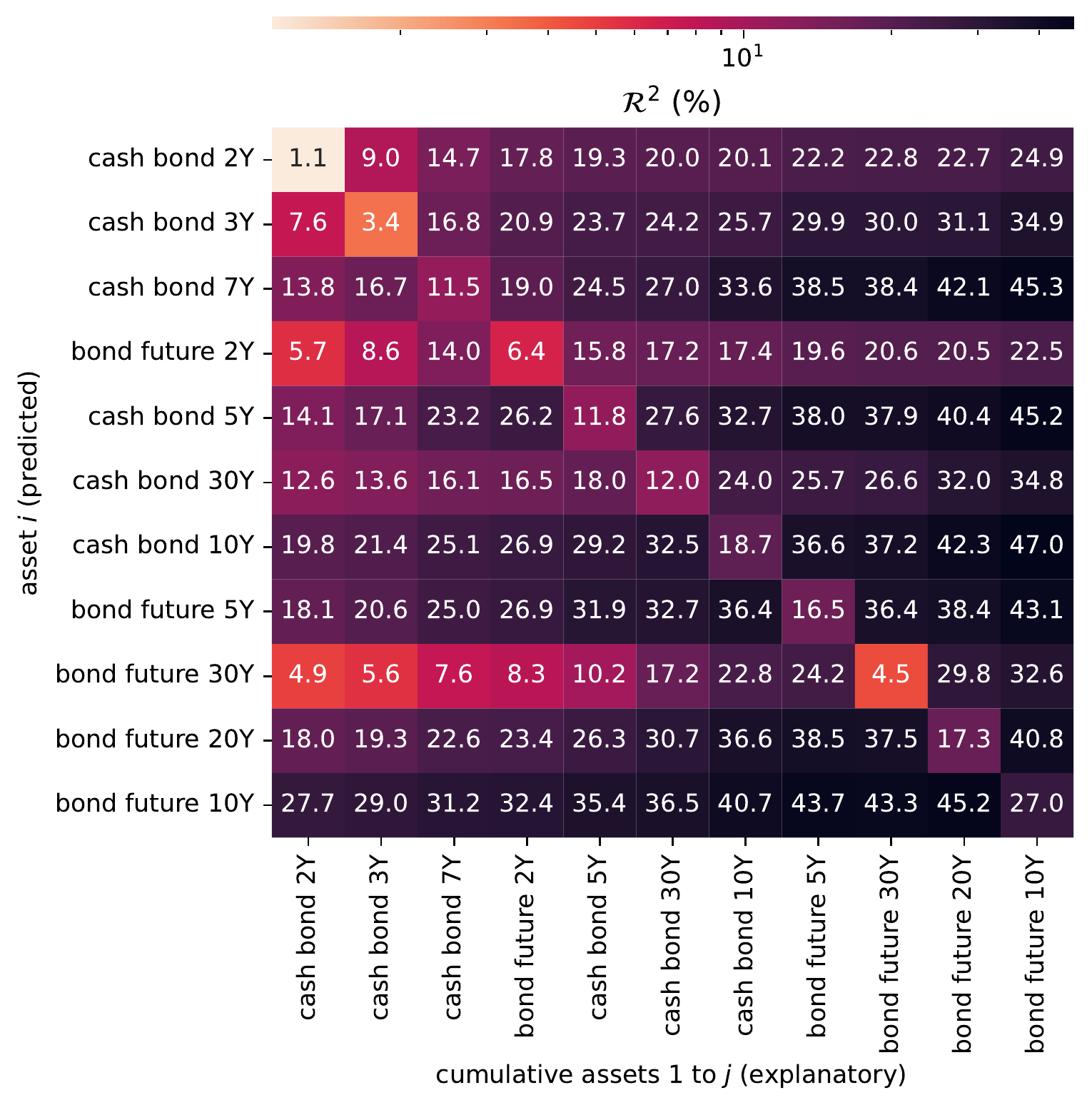}
\caption{Out-of-sample goodness-of-fit~$\mathcal{R}^2(I_{\sigma_i})$ for an increasing number of explanatory assets. The bin size was set to $30$ minutes, which is close to the optimal time scale for these assets.}
\label{fig:product_1-product_k/heatmap__R2_kyle_first-stock_out}
\end{figure}

To measure the contribution of each explanatory asset in the multidimensional case, we run the Kyle model using an increasing number of instruments. The results are presented in Fig.~\ref{fig:product_1-product_k/heatmap__R2_kyle_first-stock_out}. Each matrix item $\mathcal{R}^2_{ij}$ corresponds to the out-of-sample goodness-of-fit~$\mathcal{R}^2(I_{\sigma_i})$ regarding the prediction of asset~$i$ from the set $\{1,\dots,j\}$ of explanatory assets. In the case of diagonal items, each $\mathcal{R}^2_{ii}$ represents the effect of the diagonal model using the explanatory asset~$i$. For example, the row \textit{cash bond 5Y} can be understood as follow: its own trades explain $11.8$\% of its price increments variance, but the contribution of the other assets increase this score from $14.1$\% (using the 2-year cash bond) to $45.2$\% (using all assets).

Thus, Fig.~\ref{fig:product_1-product_k/heatmap__R2_kyle_first-stock_out} shows that we can significantly increase the explanatory power of even the most liquid asset when using a sufficiently large number of instruments of lower liquidity. More generally, this example demonstrates that a minor cross-impact effect between a pair of assets may not necessarily translate to a minor cross-impact effect at the portfolio level.

\subsection{Kyle matrix analysis}

\begin{figure} 
	\centering
	\subcaptionbox{Kyle matrix in relative price changes in basis point ($10^{-4}$ of the asset price) by $100$ millions USD worth of contract traded. \label{Lambda_kyle_relative_change}}{\includegraphics[width=\linewidth]{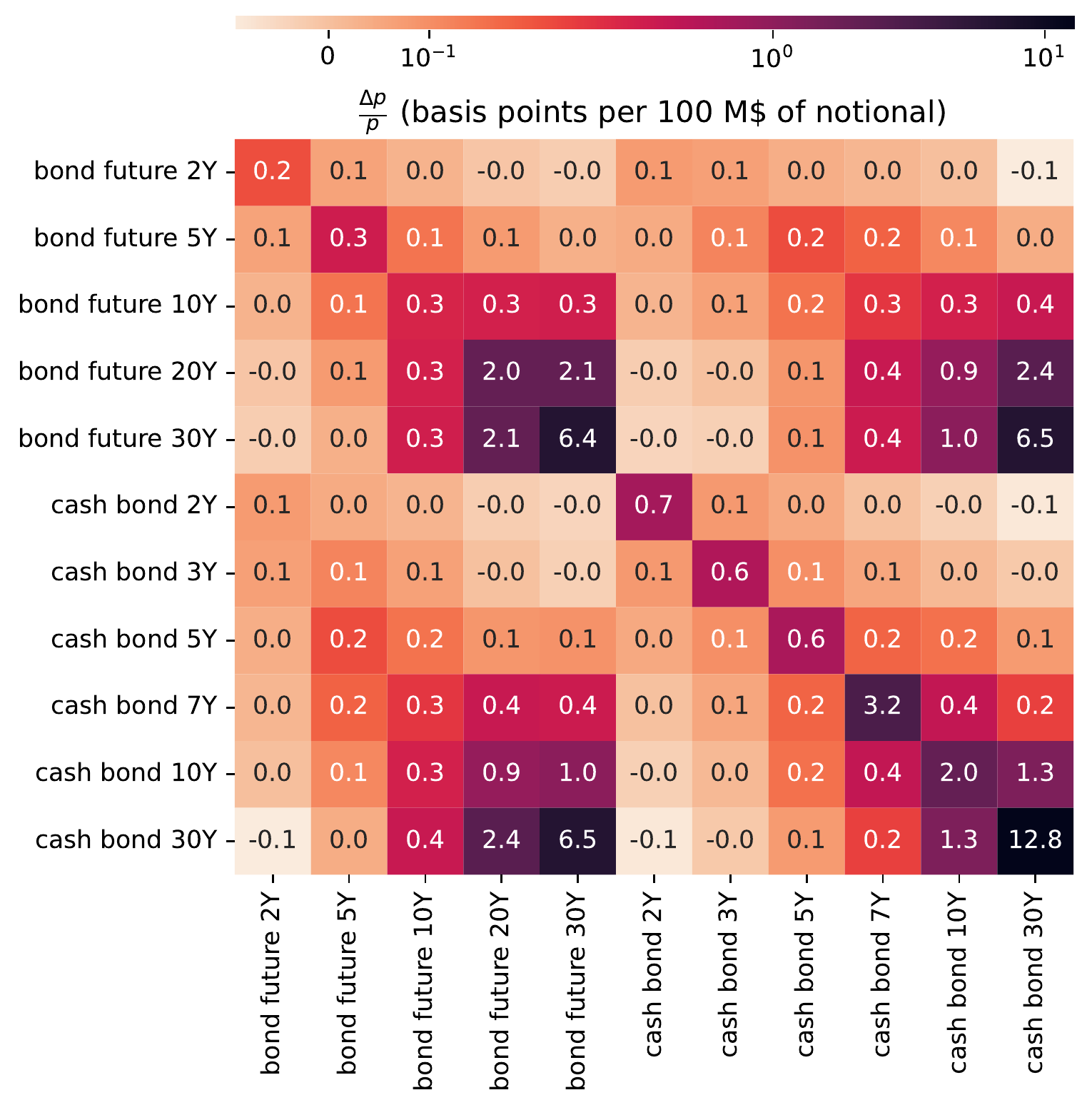}}
	\subcaptionbox{Kyle matrix in absolute variations of annual yield in basis point ($10^{-4}$ of the asset price) by $100$ millions USD worth in equivalent 10-year bond. \label{Lambda_kyle_rates}}{\includegraphics[width=\linewidth]{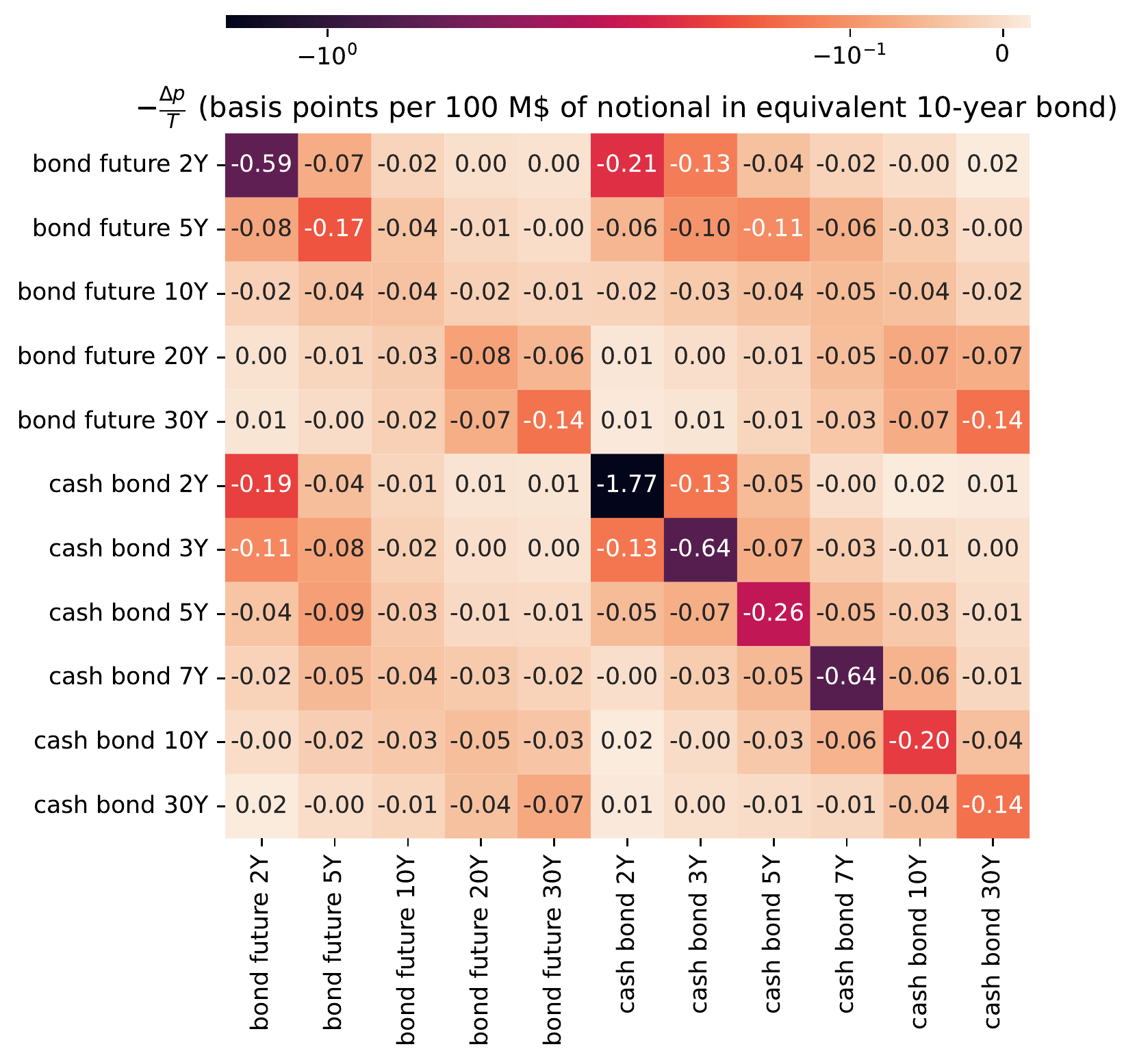}}
	\caption{Kyle matrix $\Lambda_{\text{Kyle}}$ on $9$ November $2021$ for a bin size of $30$ minutes. Units are chosen to represent either relative price changes (Fig.~\ref{Lambda_kyle_relative_change}), or absolute variations of annual yield (Fig.~\ref{Lambda_kyle_rates}).}
	\label{fig:Lambda_kyle_rates}
\end{figure}

Figure~\ref{fig:Lambda_kyle_rates} displays the Kyle matrix on $9$ November $2021$ under two different normalization conventions.

First, Fig.~\ref{Lambda_kyle_relative_change} exhibits the Kyle matrix normalized by assets' mean prices $\frac{\Lambda_{i,j}}{\bar{p}_i\bar{p}_j}$, where $\bar{p}_j = \langle p_{t,j} \rangle_t$. Thus, it defines the relative estimated price impact $\frac{\widehat{\Delta p}_{t,i}}{\bar{p}_i}$ on asset~$i$ from the traded volumes in dollars $\bar{p}_j q_{t,j}$ on asset~$j$. Indeed, one can rewrite equation~\ref{eq:prices_dynamics} as
\begin{equation}
\label{eq:prices_dynamics_relative_norm}
(\diag{\bar{p}})^{-1} \widehat{\Delta p}_t = (\diag{\bar{p}})^{-1}\Lambda_t(\diag{\bar{p}})^{-1} \diag{\bar{p}}q_t.
\end{equation}
However, this re-scaling result in over-weighting the longest tenors. Indeed, for a given interest rate $r$, the price of a zero-coupon bond contract of tenor $T$ and notional $N$ can be written as $p = \frac{N}{(1+r)^T} \approx \frac{N}{rT}$ for $r\ll1$ and $rT\gg1$. Consequently, the bond or future contract value decreases linearly with the tenor, so the normalized cross-impact matrix coefficients $\frac{\lambda_{i,j}}{\bar{p}_i\bar{p}_j}$ are proportional to the squared tenor $T^2$. This effect explains the regularities observed in Fig.~\ref{Lambda_kyle_relative_change}.

To neutralize the effect of the maturity, we propose a second normalization approach. Fig.~\ref{Lambda_kyle_rates} presents the Kyle matrix items normalized by the opposite of the tenor on the left and by the product of the price and tenor on the right: $-10\frac{\lambda_{i,j}}{T_i\bar{p}_jT_j}$, where $T_i$ is the tenor of the bond~$i$. These values define the absolute change in the interest rate $-\frac{\widehat{\Delta p}_{t,i}}{T_i}$ of the bond~$i$ from the traded volumes (in USD) in equivalent 10-year contract in the bond~$j$: $\frac{T_j}{10}\bar{p}_jq_{t,j}$. Formally, it is the reformulation of equation~\ref{eq:prices_dynamics} as
\begin{equation}
\begin{aligned}
\label{eq:rates_dynamics}
-(\diag(T))^{-1} \widehat{\Delta p}_t = \\
-(\diag(T))^{-1}\Lambda_t(\diag(T)\diag(\bar{p}))^{-1} & \diag(T)\diag(\bar{p})q_t.
\end{aligned}
\end{equation}
This second approach neutralizes the effect of the tenor on both input volumes and observed prices. Thus, the remaining differences among assets are notably due to correlation and liquidity levels. 

In this context, we are able to make four observations from Fig.~\ref{Lambda_kyle_rates}.
\begin{enumerate}
    \item Overall impacts are of similar sign and magnitude across all assets, which highlights the first factor of the interest rate curve, the \textit{parallel shift} \citep{BrigoMercurio-2006}, due to high correlations.
    \item Volumes traded on futures of a given tenor affect more significantly the interest rates of the closest tenors, which shows that correlations are higher among assets of close maturity. Equivalently, it exhibits the structure of other factor(s) beyond the parallel shift.
    \item We observe a similar behavior for cash and futures contracts, because of the high correlations between an underlying and its derivative. 
    \item As a last observation, notice that models relying on no-arbitrage in order to propagate liquidity shocks through a small number of factors (parallel shift, slope, convexity) usually predict that trading a low-liquidity asset is not expensive, as long as it is exposed to a liquid factor. In our framework, trading a low-liquidity asset is still expensive, which limits the ability to close arbitrage opportunities. This is because the empirical correlation matrices are never exactly low rank, as assumed in an idealized factor model, and heterogeneity in liquidity amplifies impact on directions of low volatility (for e.g. spread between tenor-matched cash bonds and futures). 
\end{enumerate}

\section{Conclusion}
Prices at a given time are actually influenced by the history of all previous trades through a complex process that can be formalized within the \textit{propagator model}. As its calibration is computationally intensive, our study of multidimensional price formation focuses on linear models. While the auto-correlation of signed order flows invalidates these models, they remain significant to predict prices. Notably, we have demonstrated that accurate predictions of price variations can be achieved by appropriately considering the time scale, the correlation among assets, and the liquidity, while increasing the number of explanatory assets. More importantly, we have shown that highly liquid assets determine their prices internally and that their trades influence the prices of correlated, less liquid assets. In the case of interest rate markets, the 10-year bond future serves as the main liquidity reservoir influencing the prices of the other tenors, contrary to prevailing Financial Economics theories.

However, our analysis has revealed certain gaps. Certain asset prices are best explained by their trades at significantly longer time scales than suggested by their trading frequency. More generally, our investigation into the sensitivity of optimal cross-impact time scales to asset characteristics has identified two distinct groups of asset pairs in multiple cases. Further research could explore the factors that differentiate these groups.

\section{Acknowledgments}
We would like to express our gratitude to Mehdi Tomas, Jean-Philippe Bouchaud, and Natasha Hey, who contributed to our research through fruitful discussions. We are also indebted to Bertrand Hassani, who provided us with the opportunity to conduct this study at Quant AI Lab. We extend our appreciation to Emmanuel Serie and Stephen Hardiman for their assistance in maximizing the potential of the high-performance computing tools which were instrumental in completing this study. Finally, we would like to thank Cécilia Aubrun for her helpful proofreading.

This research was conducted within the Econophysics \& Complex Systems Research Chair, under the aegis of the Fondation du Risque, the Fondation de l’École polytechnique, the École polytechnique and Capital Fund Management.

\bibliographystyle{apsrev4-2} 
\bibliography{zotero}


\appendix
\section{Notations} \label{Notations}
Table~\ref{tab:notations} summarises the notations used in this study.
\begingroup
\squeezetable
\begin{table}[h]
    \begin{ruledtabular}
    \begin{tabular}{p{0.1\linewidth} p{0.85\linewidth}}
    \multicolumn{1}{c}{Expression} & \multicolumn{1}{c}{Definition} \\
    \hline
    $n$ & The number of assets. \\
    $\mathcal{M}_n(\mathbb{R})$ & The set of real-valued square matrices of dimension $n$.\\
    $\mathcal{O}_n$ & The set of orthogonal matrices. \\
    $\mathcal{S}^{+}_n(\mathbb{R})$ & The set of real symmetric positive semi-definite matrices. \\
    $\mathcal{S}^{++}_n(\mathbb{R})$ & The set of real symmetric positive definite matrices. \\
    $A$ & A matrix. \\
    $A^\top$ & The transpose of matrix~$A$. \\
    $A^{1/2}$ & A matrix such that $A^{1/2}(A^{1/2})^\top = A$. \\
    $\sqrt{A}$ & The unique positive semi-definite symmetric matrix such that $(\sqrt{A})^2 = A$. \\
    $\diag(A) $ & The vector in $\mathbb{R}^n$ formed by the diagonal items of $A$. \\
    $\diag(v)$ & The diagonal matrix whose components are the components $(v_1, \cdots, v_n)$ of $v \in \mathbb{R}^n$.\\
    $\tau$ & The bin size. \\
    $p_{t,i}$ & The opening price of asset $i$ in the time window $[t, t + \tau ]$. \\
    $p_t$ & The vector of asset prices at opening in the time window $[t, t + \tau ]$. \\
    $q_{t,i}$ & The net market order flow traded during the time window $[t, t + \tau ]$. \\
    $q_t$ & The vector of the net traded order flows during the time window $[t, t + \tau ]$. \\
    $ \Delta p_t $ & the prices changes $p_{t+\tau} - p_t$ during the time window $[t, t + \tau ]$. \\
    $\Lambda_t$ &  The cross-impact matrix at time $t$. \\
    $\eta_t$ & The vector of zero-mean random variables representing exogenous noise at time $t$. \\
    $\Sigma_t$ & The price change covariance matrix at time $t$. \\
    $\Omega_t$ & The order flow covariance matrix at time $t$. \\
    $R_t$ & The response matrix between price variations and order flows at time $t$.\\
    $\sigma_t$ & The vector of price variation volatility at time $t$. \\
    $\omega_t$ & The vector of the signed order flow volatility at time $t$. \\
    $\mathcal{R}^2(M)$ & The $M$-weighted generalized R-squared. \\
    $\Delta \mathcal{R}^2(M)$ & The accuracy increase from the cross sectional model. \\
    $\mathcal{R}^{2*}(M)$ & The maximum goodness-of-fit observed empirically across the tested bin size~$\tau$. \\
    $\tau^*(M)$ & The optimal time scale corresponding to the maximum goodness-of-fit~$\mathcal{R}^{2*}(M)$. \\
    $\Delta \mathcal{R}^{2*}(M)$ & The maximum accuracy increase~$\Delta \mathcal{R}^2(M)$ observed empirically across the tested bin size~$\tau$. \\
    $\tau^*_{\Delta}(M)$ & The optimal time scale corresponding to the maximum accuracy increase~$\Delta \mathcal{R}^{2*}(M)$. \\
    $f_i$ & The trading frequency of the predicted asset~$i$. \\
    $f_j$ & The trading frequency of the explanatory asset~$j$. \\
    $\rho_{ij}$ & Price increments correlation between the assets~$i$ and~$j$. \\
    $\bar{\sigma}_i$ & The average across time of the price variation volatility of asset~$i$. \\
    $\bar{\omega}_i$ & The average across time of the signed order flow volatility of asset~$i$. \\
    $\bar{\omega}_i \bar{\sigma}_i$ & The liquidity of the predicted asset~$i$. \\
    $\bar{\omega}_j \bar{\sigma}_j$ & The liquidity of the explanatory asset~$j$. \\
    \end{tabular}
    \end{ruledtabular}
    \caption{Notations\label{tab:notations}}
\end{table}
\endgroup
\FloatBarrier

\section{Statistical significance analysis}
\subsection{Statistical significance of the Goodness-of-fit} \label{Statistical significance of the Goodness-of-fit}
The main results of our analysis are expressed in terms of a generalized $\mathcal{R}^{2}(M)$. For a single asset $i$, the indicator $\mathcal{R}^{2}(I_{\sigma_i})$ is precisely the R-squared of the linear regression of its price increments over its predicted price increments in the model with no Y-ratio:
\begin{equation}
\label{linear regression}
\Delta p_{t,i} = Y \widehat{\Delta p}_{t,i} + \eta_{t,i},
\end{equation}
where the explanatory variable $\widehat{\Delta p}_{t,i}$ is the prediction of the model with no Y-ratio.

The significance of the R-squared of the above regression can be provided by an F-test. Indeed, this latter allows us to compare two models, one model being the reduction of the other to fewer parameters. Here we compare the model with one explanatory variable to the model with only an intercept. Let $\widetilde{\xi} \in \mathbb{R}^N $ denote the vector of the errors estimated in the model with no explanatory variable, and $\widehat{\xi} \in \mathbb{R}^N$ denote the vector of the errors estimated in the cross-impact model. The F-statistic is expressed as the normalized difference between the squared errors in the two models:
\begin{equation}
    F = \frac{\widetilde{\xi}^\top \widetilde{\xi}-\widehat{\xi}^\top \widehat{\xi} } { \widehat{\xi}^\top \widehat{\xi}}.
\end{equation}
Since the errors in the parameter-free model are precisely equal to the centered explained variable, denoted as $y \in \mathbb{R}^N$, we can express this F-statistics as a function of the $\mathcal{R}^{2}$:
\begin{equation}
    F = \frac{y^\top y - \widehat{\xi}^\top \widehat{\xi}} {\widehat{\xi}^\top \widehat{\xi}} = \frac{\widehat{y}^\top \widehat{y}}{\widehat{\xi}^\top \widehat{\xi}} = \frac{\mathcal{R}^{2}}{1-\mathcal{R}^{2}}.
\end{equation}
Under the usual assumptions for the residuals and the explained variable, this F-statistics follows a Fisher law $F(N,N-1)$. The significance of the $\mathcal{R}^{2}$ can be then provided by the p-values of the F-statistics of the linear model calibrating the Y-ratio.

Yet, the residuals in the above regression are auto-correlated due to the properties of the order flows. They are also non-Gaussian (heavy tails, negative skew) and heteroskedastic, due to the properties for the price process. In fact, returns are generally conditionally heteroskedastic but unconditionally homoskedastic. Here the unconditional heteroskedasticity observed in the data might be due to some trend in the annual sample.

However, the observed levels of auto-correlation are sufficiently low (around $10\%$ at the lag $1$) to avoid compromising the robustness of the test \citep{Kramer-1989}. This issue is further studied in the following section. Moreover, the non-Gaussianity only partially limits the robustness of the F-statistic test \citep{BoxWatson-1962}. Yet, the heteroskedasticity issue requires using a modified F-statistic test robust to this assumption. Thus, we measure the statistical significance of the $\mathcal{R}^{2}$ using the approach of \citet{MacKinnonWhite-1985} (implemented in the \textit{Statsmodels} python library through the method of \citet{LongErvin-2000}).

\begin{figure}
	\centering
\includegraphics[width=\linewidth]{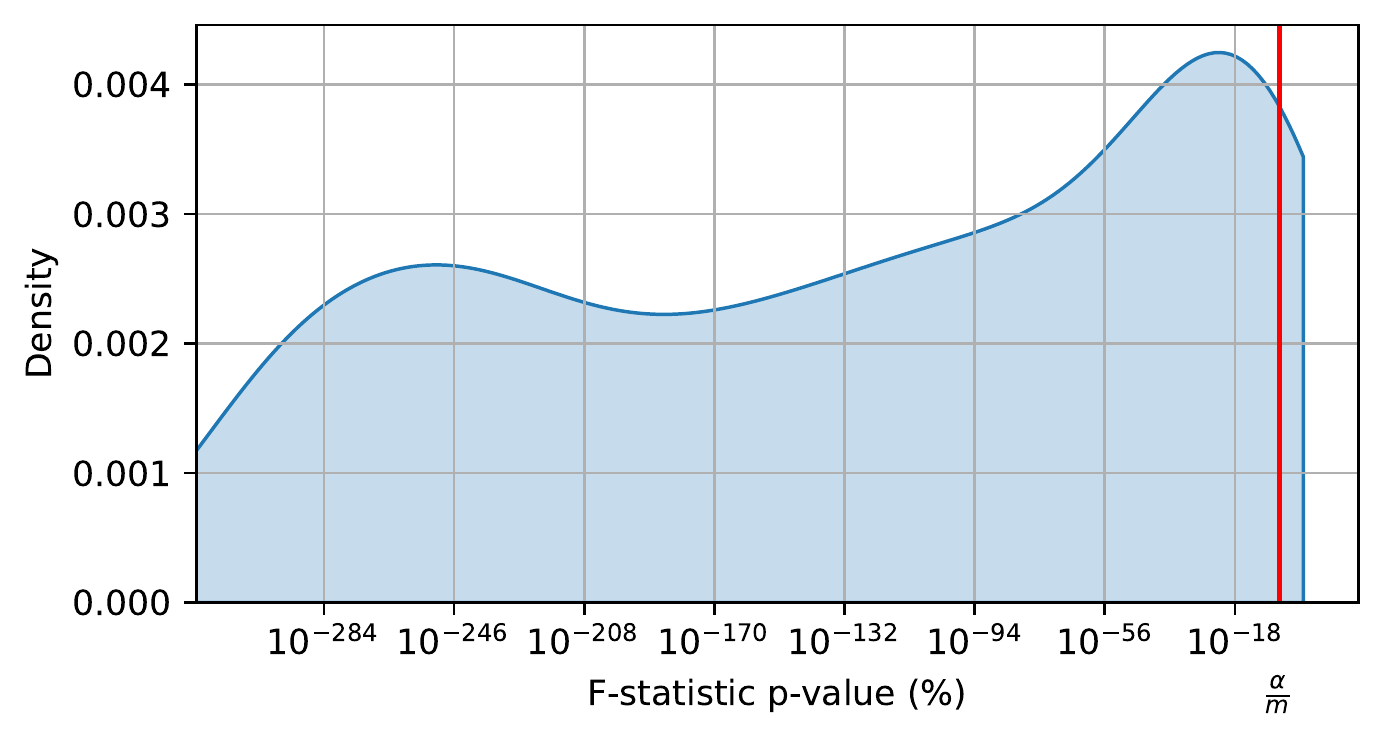}
	\caption{Empirical distribution of the statistical significance of the $\mathcal{R}^{2*}(I_{\sigma_i})$ across the years and assets in our sample.}
\label{fig:heatmap_F_p-value_R2_kyle_first}
\end{figure}

These F-statistics confirm that the $\mathcal{R}^{2*}$ displayed in our study are significant. Specifically, in the single asset case, each optimal goodness-of-fit $\mathcal{R}^{2*}$ is obtained from a linear regression. The F-statistics p-values of $m\approx10^3$ ($500$ assets across $5$ years) statistical tests are exhibited in Fig.~\ref{fig:heatmap_F_p-value_R2_kyle_first}. Notably, only few p-values are above the Bonferroni upper bound \citep{GoemanSolari-2014,Frane-2015}. This upper bound is used for the identification of false positive when performing multiple hypothesis tests. Here, the rate of false positive at the confidence interval $\alpha =10^{-2}$ is bounded by the share of p-values above $\frac{\alpha}{m}=10^{-5}$. Figure~\ref{fig:heatmap_F_p-value_R2_kyle_first} shows that only a negligible share of these p-values are not significant (approximately $2$\%). The more accurate procedure from \citet{BenjaminiHochberg-1995} yields similar results. If we compare the p-value of rank $k$ (in ascending order) to $\frac{k\alpha}{m}$, we find that $1.8$\% of these p-values are above this threshold.

\FloatBarrier

\subsection{Auto-correlation structure and comparison with the propagator model} \label{Auto-correlation structure and comparison with the propagator model}
\begin{figure}
	\centering
\includegraphics[width=\linewidth]{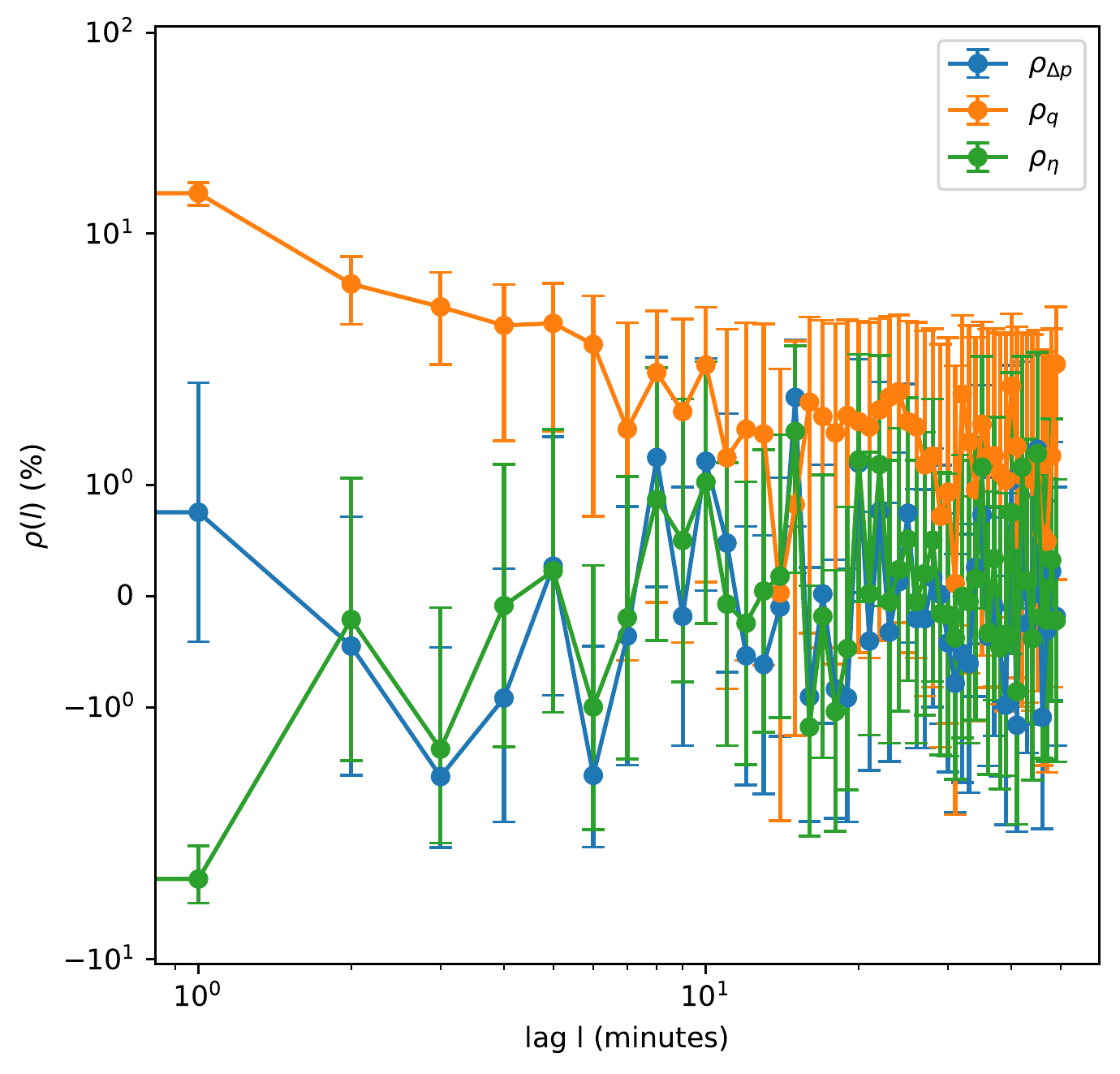}
	\caption{Auto-correlation of price variations $\Delta p$, signed order flows $q$ and residuals $\eta$ for the E-mini S\&P future. Data is binned every $1$ minute for the year $2021$. Error bars represent one standard deviation confidence interval.}
\label{fig:auto-correlations}
\end{figure}

The auto-correlation of signed order flows is a well-documented feature of financial markets \citep{LilloFarmer-2004}. Using the E-mini S\&P future binned every $1$ minute to calibrate the single asset model, we observe significant auto-correlation of both the signed order flows and the residuals (Fig.~\ref{fig:auto-correlations}). In contrast, the auto-correlation of prices is of the same size as the noise. If the number of data points increases, prices will become even more efficient, so the auto-correlation of the residuals will increase to compensate for the long memory of the signed order flows. Thus, the model will be invalidated.

As previously mentioned, one approach to re-conciliate the long memory of the order flows with the efficiency of prices is to define a \textit{propagator model} \citep{BouchaudEtAl-2006,Bouchaud-2009, AlfonsiEtAl-2016, BenzaquenEtAl-2017, BouchaudEtAl-2018, SchneiderLillo-2019} as follow:
\begin{equation}
    p_t = \sum_{s \leq t} G(t-s)q_s + \eta_t,
\end{equation}
where $G:t\to G(t) \in \mathcal{M}_n(\mathbb{R}) $ captures the dependence on past order flows and $\eta_t$ is a vector of zero-mean random variables. As shown by \citet{TomasEtAl-2022} the calibration of the true propagator model would yield only marginal improvements in the goodness-of-fit. However, this model is significantly more complex to calibrate, which would impede conducting this study at the same scale across time and assets.

\end{document}